\documentclass[lettersize,journal]{IEEEtran}
\usepackage{amsmath,amsfonts}
\usepackage{array}
\usepackage[caption=false]{subfig}
\usepackage{textcomp}
\usepackage{stfloats}
\usepackage{url}
\usepackage{verbatim}

\usepackage{graphicx}
\usepackage{cite}
\usepackage{pifont}
\usepackage[noend]{algpseudocode}
\usepackage{makecell}

\usepackage{amsmath}
\usepackage{amssymb}
\usepackage{booktabs}
\usepackage{multirow}
\usepackage[pagebackref,breaklinks,colorlinks]{hyperref}
\usepackage[table]{xcolor}
\usepackage{xcolor}
\definecolor{r}{rgb}{1,0,0}

\begin{document}
\title{Omni-C: Compressing Heterogeneous Modalities into a Single Dense Encoder}

\author{Kin Wai Lau$^{1,2}$, Yasar Abbas Ur Rehman$^{2}$, Lai-Man Po$^{1}$, Pedro Porto Buarque de Gusmão$^{3}$ \\
City University of Hong Kong$^{1}$\\
TCL AI Lab$^{2}$\\
University of Surrey, United Kingdom$^{3}$ \\
}

\markboth{Journal of \LaTeX\ Class Files,~Vol.~XX, No.~XX, Feb~2026}%
{Shell \MakeLowercase{\textit{et al.}}: A Sample Article Using IEEEtran.cls for IEEE Journals}


\maketitle

\begin{abstract} 
Recent multimodal systems often rely on separate expert modality encoders which cause linearly scaling complexity and computational overhead with added modalities. While unified Omni-models address this via Mixture-of-Experts (MoE) architectures with specialized experts and routing, they still inflate parameter counts and introduce routing overhead. In this paper, we propose Omni-C (Omni-Compress), a single dense Transformer-based encoder that learns competitive shared representations across heterogeneous modalities—images, audio, and text—through unimodal contrastive pretraining on large-scale unaligned data. By maximizing parameter sharing in the backbone and using lightweight modality-specific projection heads, Omni-C effectively mitigates inter-modality conflicts without requiring MoE, paired supervision, or routing. This design supports efficient deployment on memory-constrained systems via sequential modality processing and low-memory inference, eliminating the need for parallel expert loading or specialized hardware. Experiments show Omni-C achieves performance comparable to expert models in unimodal and cross-modal tasks, with modest zero-shot degradation on audio and text that is largely recovered through lightweight linear probing or parameter efficient fine-tuning. The unified architecture substantially reduces inference memory usage compared to multi-encoder baselines, advancing efficient and scalable multimodal learning. Code are available at \url{https://github.com/StevenLauHKHK/Omni-C}.
\end{abstract}


\section{Introduction}
Learning universal feature representations via Self-Supervised Learning (SSL) across multiple modalities has gained significant traction over the past decade. Current trends in unimodal and multimodal understanding heavily rely on expert encoders—large foundational models pretrained on vast amounts of images, audio, or text. The availability of these specialized encoders has accelerated the development of multimodal systems for tasks such as intra-modal and cross-modal recognition, retrieval, segmentation, and zero-shot inference \cite{girdhar2023imagebind, guzhov2022audioclip, li2025uni, zhang2025assessing}. However, incorporating each new modality-specific encoder substantially increases system complexity, especially when architectures differ in computational requirements and processing pipelines. 

Existing work on unified encoders falls into two main paradigms:
(1) unifying sub-modalities within a single domain (e.g., images, videos, depth, 3D) via a shared backbone \cite{girdhar2022omnivore};
(2) integrating heterogeneous modalities (e.g., vision, audio, text) through specialized encoders combined with fusion or alignment layers (such as ImageBind and Meta Transformer) \cite{girdhar2023imagebind, zhang2023meta, li2025uni}, and gating layers (such as MoE) \cite{wu2024omni, li2025uni, li2025uni2, ai2025ming, xu2025qwen3omnitechnicalreport}. Despite these advances, training a single, end-to-end unified model (single model) across truly diverse modalities—such as images, audio, and text—remains underexplored, particularly in terms of balancing parameter efficiency, cross-modal knowledge transfer, and preservation of strong unimodal performance.

\begin{figure}
   \centering
   \includegraphics[width=\linewidth]{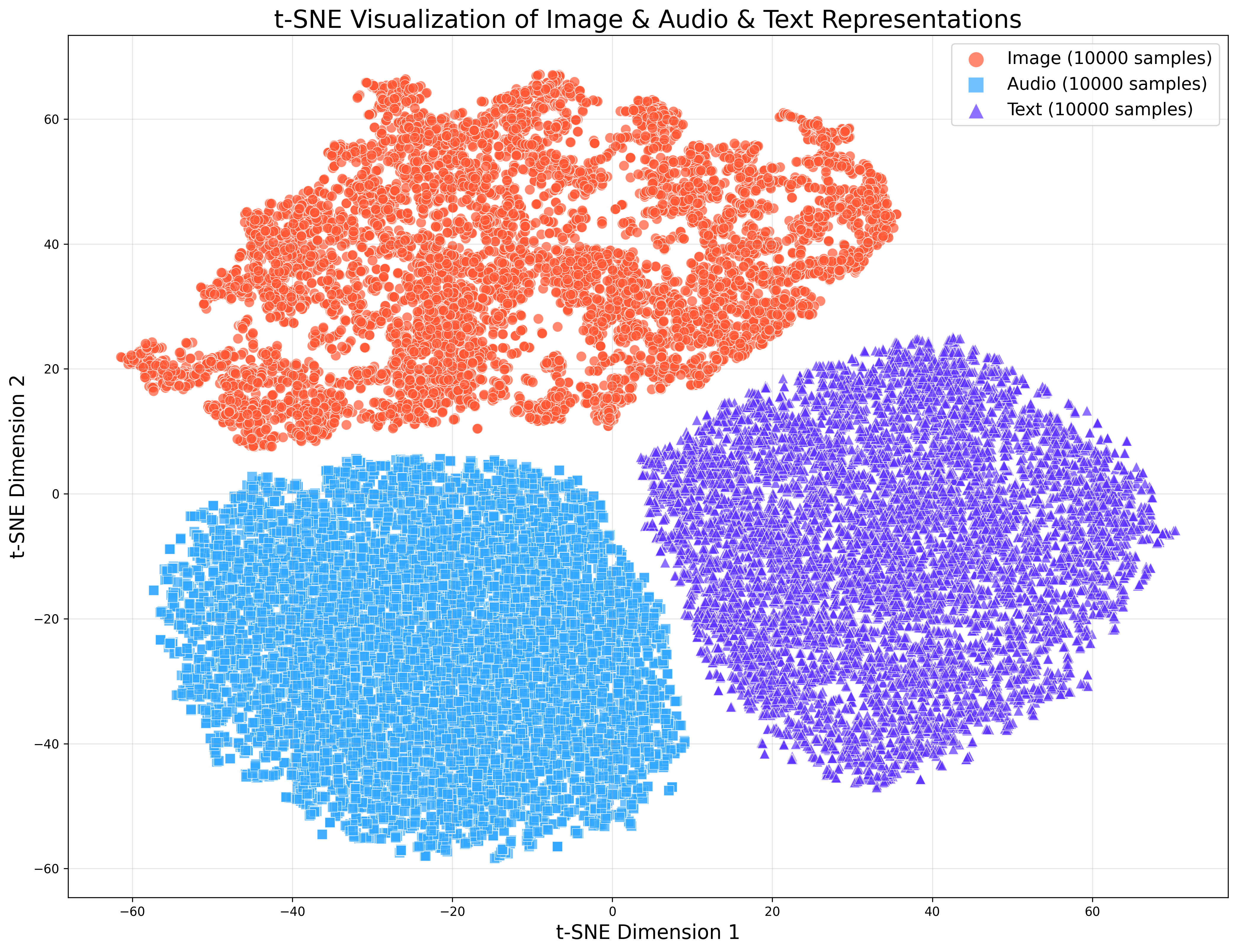} 
   \caption{t-SNE visualization of image, audio, and text embeddings from pretrained Omni-C model. It shows clear separation of image (red), audio (blue) and text (green) clusters. Embeddings are extracted from samples on ImageNet-1K (images), AudioSet (audio spectrograms), and English Wikipedia (text)}
   \label{fig:tsne_image_audio_text_unified_seperate_projector}
\end{figure}

Motivated by these limitations, we pose the following research question: 

\noindent \textit{\textbf{Can a single unified encoder, trained jointly on audio, visual, and text modalities, achieve competitive or comparable performance to expert counterparts without relying on explicit gating or routing mechanisms?}}

Addressing this research question yields two primary benefits.
(1) The resulting unified encoder provides a shared representation space capable of approximating heterogeneous modalities (images, audio, and text) in a joint embedding, facilitating the learning of rich transferable feature representations. These approximate features can then be further refined via supervised fine-tuning for improved performance on specific downstream tasks.
(2) Training a single model to process multiple modalities reduces overall system complexity compared to maintaining separate modality-specific models, as the addition of new modalities no longer requires the development and integration of entirely new encoders.

 In this work, we pursue this direction by developing Omni-C (Omni-Compress), a unified encoder that acts as a powerful lossy compressor for heterogeneous modalities and is jointly pretrained on images, audio spectrograms, and text. A critical criterion for the success of this approach is that the learned representations form distinct global clusters on the hypersphere for different modalities (see Fig. \ref{fig:tsne_image_audio_text_unified_seperate_projector}), while within each modality maintaining discriminative power through high similarity of positive pairs and low similarity of negative pairs \cite{gao2021simcse, chen2020simple}. To achieve this, we adapt the sequential training strategy from Omnivore \cite{girdhar2022omnivore} and pretrain Omni-C on images, audio, and text using SSL contrastive learning \cite{chen2020simple}. We opt for SSL pretraining over supervised alternatives in Omnivore for its practicality: It enables effective utilization of both labeled and large-scale unlabeled datasets. Moreover, SSL is better suited to our goal of learning global, transferable feature representations across modalities. In particular, training a single encoder using SSL eliminates the need for paired data, allowing seamless leverage of abundant unpaired and unlabeled multimodal corpora \cite{geng2022multimodal}. 

Another essential criterion for Omni-C to act as lossy compressor is that the share backbone naturally develops distributed attention patterns across input patches when jointly pretrained on heterogeneous modalities (see Fig. \ref{fig:attn-map-omni-image}, \ref{fig:attn-map-omni-audio}, and \ref{fig:attn-map-omni-text}). This allows the model to concurrently encode and represent information from multiple modalities in a shared space. In contrast, unimodal expert models tend to exhibit focused attention (see Fig.\ref{fig:attn-map-expert-image}, \ref{fig:attn-map-expert-audio}, and \ref{fig:attn-map-expert-text}), specializing in modality-specific features. This distinction echoes well-established concepts in perceptual psychology, where distributed attention enables rapid extraction of global, gist-like scene summaries, while focused attention supports precise identification of individual elements \cite{10.1093/acprof:oso/9780199658442.003.0002}. Although the transformer literature \cite{vaswani2017attention, dosovitskiy2020image} rarely frames attention in these exact psychological terms, our findings suggest that a shared backbone encourages distributed attention patterns conducive to holistic, cross-modal representations -- potentially enhancing transferability and efficiency without explicit modality silos. This insight underpins our design of Omni-C, which leverages such emergent properties to achieve competitive multimodal performance with a simplified, single-encoder architecture.

To evaluate the effectiveness of this lossy universal compressor, we first pretrain Omni-C on audio, image, and text using SSL contrastive learning. We then assess the pretrained model on diverse downstream tasks, including zero-shot inference, linear-probing, low-rank adaptation \cite{po2024sbora}, and cross-modal alignment. Results demonstrate that the unified model delivers competitive performance across modalities while maintaining a simplified architecture.

Our contributions can be summarized as follow:
\begin{itemize}
    \item We propose Omni-C (Omni-Compress), a unified dense encoder for multimodalities that eliminates the need for parallel expert loading and MoE routing, significantly reducing inference memory usage.
    \item We validate Omni-C as a lossy universal compressor that produces robust global representations, which can be effectively restored via parameter-efficient fine-tuning such as SBoRA \cite{po2024sbora}.
    \item We demonstrate effective cross-modal alignment using a linear-probe approach inspired by SAIL\cite{zhang2025assessing}, achieving competitive cross-modal zero-shot performance.
    \item We solve inter-modality feature conflicts through the strategic use of modality-specific projection heads, ensuring clear separation of modalities in the shared embedding space.
\end{itemize}

The rest of this paper is organized as follows: Section \ref{sec:literature} discusses the related work on unified models for multiple modalities and multi-modal alignment. Section \ref{sec:methodology} illustrates the methodology for training a single model with multiple heterogeneous modalities. Section \ref{sec:experiments} provides experimental results and analysis. Section \ref{sec:ablation_studies} dissects the impact of our design choices of our Omni-C. We concluded the paper with a conclusion in Section \ref{sec:conclusion}.

\section{Related Work}
\label{sec:literature}
\subsection{Unified Models for Multiple Modalities}

Recent advances have explored training single models that handle multiple modalities to achieve shared representations and reduce system complexity. Omnivore \cite{girdhar2022omnivore} proposes a unified ViT for labelled visual modalities (RGB images, videos, single-view RGB-D) that achieves parameter sharing via spatio-temporal patching and joint supervised training on unaligned classification tasks, yielding strong cross-modal generalization. Similarly, OmniVLA \cite{hirose2025omnivla} extends unified multimodal architectures to physical robotic manipulation by integrating infrared, mmWave, and acoustic modalities. It introduces sensor-masked images as a unified representation by overlaying spatial grounded sensor data onto RGB images to enable data-efficient fine-tuning of an RGB-pretrained Vision-Language-Action backbone (VLA) with lightweight per-sensor projectors. ImageBind \cite{girdhar2023imagebind} binds six modalities (images, text, audio, depth, thermal, IMU) using images as an anchor to pair heterogeneous data, enabling emergent zero-shot alignment without all-pair supervision. However, these methods still require paired or labelled data to bind the modalities together, and the approaches like OmniVLA involve additional alignment processes for overlaying sensor data and extra segmentation modules.

More recent efforts incorporate Mixture of Experts (MoE) for scaling unified multimodal models. Omni-SMoLA \cite{wu2024omni} uses soft MoE with low-rank experts for vision-language tasks to boost generalist performance. Uni-MoE \cite{li2025uni} and its extension Uni-MoE-2.0-Omni \cite{li2025uni2} employ MoE to efficiently handle diverse modalities including text, image, audio and video. These models incorporate modality-specific routing mechanisms and progressive training strategies to manage multimodal inputs while maintaining scalability and generalization. Similarly, Ming-Omni \cite{ai2025ming} and Qwen3-Omni \cite{xu2025qwen3omnitechnicalreport} leverage MoE for real-time omnimodal processing across speech, vision, and text. However, MoE-based methods often introduce routing overhead, higher training complexity due to expert balancing, and increased memory demands during inference from sparse expert activation.

Another line of work, known as multimodal pathway approaches \cite{zhang2024multimodal}, improves the unimodal transformers by injecting irrelevant data from other modalities. This method enhances a target transformer using an auxiliary transformer trained on a different modality. For this purpose, it builds the neural pathways by adding auxiliary weights as parallel linear branches during training. The auxiliary weights are merged into the target branch via reparameterization during inference. However, this multi-pathway design relies on parallel branches to process target and auxiliary models during training. As the number of auxiliary modalities or models increases, both computational usage and memory footprint grow linearly due to simultaneous handling of multiple transformers.

In contrast to these works, our approach trains a single dense model on heterogeneous modalities, including images, audio spectrograms, and text, using unimodal contrastive learning on unaligned datasets.  This design achieves a simpler, more efficient lossy compressor that competes with experts across a range of downstream tasks.

\subsection{Vision/Audio-Language Alignment Methods}
Vision-language alignment leverages contrastive training on large paired datasets. For instance, CLIP \cite{radford2021learning} jointly trains image and text encoders from scratch on 400 million pairs using InfoNCE loss for zero-shot transfer. ALIGN \cite{jia2021scaling} scales it to 1.8 billion noisy pairs for improved robustness. Florence-VL \cite{chen2025florence} improves the vision-language models by leveraging Florence-2's generative vision encoder \cite{xiao2024florence} and a novel Depth-Breadth Fusion architecture to integrate hierarchical vision features and task-specific features into the pretrained Large Language Models (LLMs). Through the end-to-end pretraining and target fine-tuning on diverse datasets, it achieves state-of-the-art performance on various benchmarks like Visual Question-Answers (VQA), Opitcal Character Recognition (OCR), and object hallucination tasks. Similarly for the Audio-language alignment, CLAP \cite{elizalde2023clap} applies CLIP-style contrastive learning to audio-text pairs to achieve strong zero-shot performance on audio tasks. AudioCLIP \cite{guzhov2022audioclip} extends CLAP by training on audio-image-text triplets for zero-shot classification and retrieval. More recent works like SALMONN \cite{tang2023salmonn} achieve audio-text alignment by integrating speech and audio encoders with a pre-trained text-based LLM through a Q-Former connection module and LoRA adapters trained on multimodal datasets. These methods typically require end-to-end pretraining of the full model or major components on paired data to establish effective cross-modal alignment, which incurs high computational cost. 

Some recent approaches like LiT \cite{zhai2022lit} reduce this burden by freezing a pretrained vision encoder and tuning only the text encoder on image-text pairs thereby drastically lowering computation while maintaining competitive zero-shot transfer performance. More recent works like SAIL \cite{zhang2025assessing} align frozen vision and language backbones with lightweight linear or non-linear layers on limited paired data. The training pairs are only 6\% of CLIP's scale. SAIL employs a refined sigmoid-based contrastive loss for better hard-negative handling. As a results, SAIL reduces compute requirements while matching CLIP's zero-shot performance on retrieval and classification tasks. 

Our method extends this line of work by leveraging our Omni-C model, which is pretrained with embedded multimodal knowledge from unaligned sources, as the unified backbone for alignment. In contrast to approaches that rely on separate expert encoders for each modality, such as those in CLIP or SAIL, we use a single shared model. This design avoids the need for multiple backbones and achieves comparable alignment performance with greater training and inference efficiency.

\begin{figure*}[t]
\centering
\includegraphics[width=0.95\linewidth]{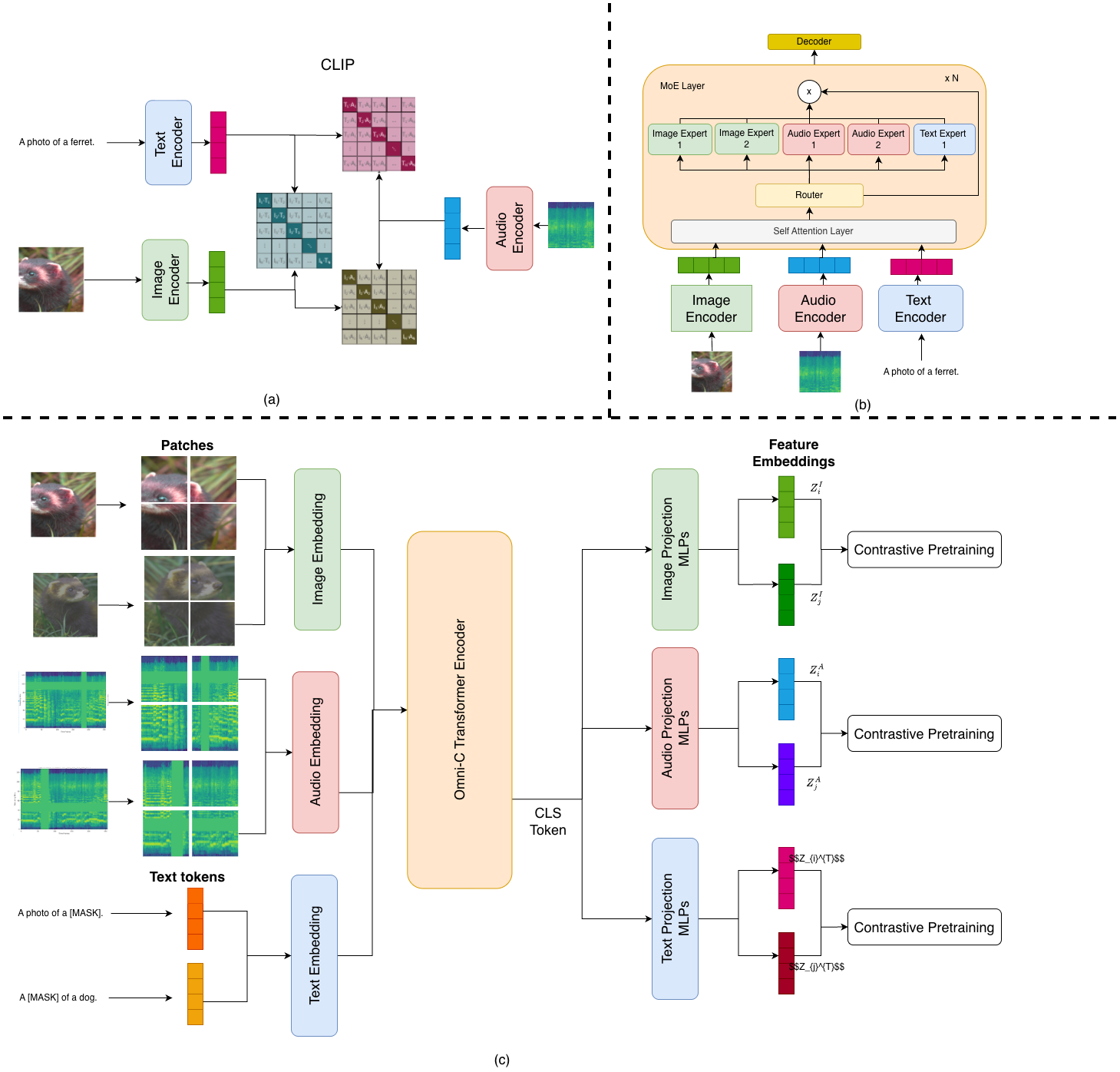}
\caption{Unlike the multi-expert (e.g. AudioCLIP \cite{guzhov2022audioclip}) in (a) and Mixture-of-Experts (MoE) approaches (e.g. , Uni-MoE 2.0-Omni \cite{li2025uni2}) in (b) which incur linear parameter scaling and routing overhead with added  modalities, Omni-C in (c) leverages a single dense Transformer backbone with maximal parameter sharing to achieve competitive unimodal and cross-modal performance while drastically reducing system complexity and inference memory during the deployment. Omni-C model processing multiple heterogeneous modalities (images, audio spectrograms, and text). Images and audio spectrograms are divided into non-overlapping patches and projected via separate 2D convolutional embedding layers, while text sequences are tokenized and projected via a linear embedding layer. A shared learnable global CLS token is prepended to the sequence of embeddings (with modality-specific positional encodings added), and the full sequence is processed by the unified Omni-C Transformer backbone blocks. The final CLS token representation from the backbone is then fed into modality-specific MLP projection heads for unimodal contrastive pretraining.}
\label{fig:Omni-C}
\end{figure*}

\section{Methodology}
\label{sec:methodology}
Our goal is to learn a unified Omni-C (Omni-Centralized) model that can operate on three different common modalities, including images, audio, and text. However, each input modality to the Omni-C has different dimensions and sizes. For instance, an image has three channels (RGB), an audio spectrogram has only one channel, and text has one dimension without width and height. Therefore, it is required to convert all the inputs to the same embedding dimension before feeding them into the unified backbone. To tackle this problem, we follow the approach of \cite{dosovitskiy2020image, devlin2019bert, gong2022ssast} and adopt two independent convolution patch embedding layers for image and audio, and one linear layer for text encodings. We adopt the ViT architecture as the backbone model because its self-attention mechanism gracefully handles variable-sized inputs for different modalities. Figure \ref{fig:Omni-C} presents an overview of our approach. 

\subsection{The Omni-C Model (Image \& Audio \& Text Modality)}
We propose a unified Omni-C transformer-based architecture that jointly processes image, audio spectrogram, and text modalities, emphasizing maximal parameter sharing in the core backbone. Inspired by Omnivore \cite{girdhar2022omnivore}, our model learns general-purpose representations from large-scale unaligned and unlabeled data across these heterogeneous modalities using contrastive SSL. 

\noindent\textbf{Input representations.} Omni-C accepts three input formats tailored to each modality: RGB images $I \in \mathbb{R}^{3 \times H \times W}$, where $H$ and $W$ denote the height and width of the image. Audio spectrograms $S \in \mathbb{R}^{1 \times H' \times W'}$ are log-mel spectrograms treated as a single-channel image, where $H'$ represents the number of frequency bins (vertical axis) and $W'$ represents the time axis (horizontal axis). Text sequences are tokenized using a BERT tokenizer \cite{devlin2019bert} to produce integer token IDs $T \in \mathbb{R}^{L}$, where $L$ denotes the fixed sequence length.

\noindent\textbf{Patch and token embeddings.}  
Following the standard approach from prior studies \cite{dosovitskiy2020image, gong2022ssast}, we first convert images and audio log-mel spectrograms into $d$-dimensional patch embeddings. Specifically, we pass image and log-mel spectrograms through a dedicated 2D convolutional layer with output channels equal to the embedding dimension $d$. The convolutional kernel $(h \times w)$  matches the stride size, ensuring non-overlapping patches and converting an input of dimensions $H\times W$ into a feature map of size $(H/h)\times(W/w)\times d$. This feature map is then flattened along the spatial (width-height) dimension to yield the final embeddings $d\times ps$, where $ps = (H/h)\times(W/w)$ denotes the total number of patches. For text inputs, we use the BERT \cite{devlin2019bert} to tokenize the text into an integer sequence $T \in \mathbb{R}^{L}$, which is then mapped to the same embedding dimension $d$ through a learnable projection layer to get the token embeddings.

\noindent\textbf{Unified backbone and positional encoding.}  We adopt a Vision Transformer (ViT) encoder\cite{dosovitskiy2020image}, denoted as $f$, to map the patch or token embeddings from the three modalities into a shared representation space. Prior to feeding these embeddings into the encoder, we add modality-specific positional embeddings: 2D sinusoidal positional embeddings for each image and audio patch token, and 1D sinusoidal positional embeddings for each text token. Additionally, following the standard ViT approach\cite{dosovitskiy2020image}, we prepend a learnable CLS token to each input sequence to capture global context. The resulting sequences, comprising the CLS tokens and positional-augmented embedding tokens, are then processed through the ViT blocks. Finally, the output CLS tokens are passed to the projection layers.

\noindent\textbf{Projection heads.} To facilitate unimodel contrastive pretraining \cite{chen2020simple, gao2021simcse, rehman2025fssuavl} on unaligned data, the CLS token representation from the shared backbone is fed into the modality-specific multi-layer perceptron (MLPs) projection heads. Each head consists of a two-layer MLP, which is a linear transformation followed by RELU activation and a final linear layer. These MLPs project the representation into a lower-dimensional space optimized for the contrastive objective while retaining the advantages of a unified encoder.

This design enables the model to acquire robust representations across image, audio, and text modalities solely from unaligned unimodality, similar to the design of Omnivore \cite{girdhar2022omnivore}.

\subsection{Pretraining the Omni-C Model (Image-Audio-Text Modality)}
Omni-C produces a unified embedding  $F(X) = \Phi$ for multiple modalities, i.e., image, audio spectrogram, and text. Unlike supervised multi-task learning approaches \cite{girdhar2022omnivore} \cite{girdhar2023imagebind} that rely on shared label spaces or aligned data, we pretrain our model in a fully self-supervised manner using large-scale unaligned datasets with no cross-modal correspondences and no overlapping supervision signals.

We jointly pretrain Omni-C on three independent unimodal datasets: ImageNet-1K\cite{deng2009imagenet}, AudioSet\cite{gemmeke2017audio}, and the English Wikipedia corpus\cite{coster-kauchak-2011-simple}. This setup draws inspiration from the multi-dataset training paradigm of Omnivore\cite{girdhar2022omnivore}. However, our approach differs fundamentally in terms of supervision and the incorporation of heterogeneous modalities. Instead of using dataset-specific classification heads with supervised cross-entropy losses, we employ unimodal contrastive learning~\cite{chen2020simple, gao2021simcse, rehman2025fssuavl} independently within each modality.

In each training iteration, following Omnivore\cite{girdhar2022omnivore}, we employ a modality-separated minibatch strategy. We sample a modality and construct a minibatch consisting solely of samples from that modality. Within this minibatch, two randomly augmented views are generated for each input sample\cite{chen2020simple, gao2021simcse} that is passed through the shared backbone to produce feature representations. The CLS token of the feature representations is then passed through modality-specific MLP projection heads to produce lower-dimensional projected embeddings. Finally, we compute the contrastive loss~\cite{chen2020simple} between these embeddings, i.e.,  pulling the representations of the two augmented views of the same sample closer together while pushing apart those of different samples in the batch.

This modality-separated minibatch strategy \cite{girdhar2022omnivore}, along with the use of separate modality-specific projection heads, ensures stable training, resulting in distinct clusters of each modality feature representations that are spread uniformly on the hypershpere. In contrast, employing a single shared projection head across all modalities results in a shared embedding space where the modalities representations are mixed together resulting in performance degradation. We examine this limitation in ablation studies in Section \ref{sec:mp-vs-sp}.

\section{Experiments}
\label{sec:experiments}
We conducted a comprehensive set of experiments to evaluate the effectiveness of our Omni-C unified encoder. We compare our pretrained model against modality-specific baselines across a range of downstream tasks evaluations, including zero-shot evaluation, linear probing for classification, Parameter-Efficient Fine-Tuning (PEFT), and multimodal alignment. In addition, we also performed ablation studies to investigate key design choices, such as projection head configurations and the model's alignment and uniformity properties in the embedding space. To further assess the impact of modality mixing in the unified backbone, we also provide a set of variant Omni-C models trained with different modality combinations such as image-text only, audio-text only and image-audio only. 

\noindent\textbf{Pretraining datasets.} We pretrain our model on three large-scale, unaligned unimodal datasets:
\begin{itemize}
    \item \textbf{Images:} ImageNet-1K \cite{deng2009imagenet} contains approximately 1.28 million training images and 50K validation images across 1000 object categories.
    \item \textbf{Audio:} Audioset \cite{gemmeke2017audio} contains over 2 million 10-seond YouTube audio clips from the training split, annotated with 632 audio event classes. Each clip is converted to a fix-size log-mel spectrogram.
    \item \textbf{Text:} The English Wikipedia corpus \cite{coster-kauchak-2011-simple} provides a diverse collection of articles yielding billions of tokens after BERT tokenization \cite{devlin2019bert}.
\end{itemize}

\noindent\textbf{Pretraining Implementation details.} We employ a standard Vision Transformer (ViT) \cite{dosovitskiy2020image} as our shared backbone in the Base configuration with patch size 32 (ViT-B/32), which includes 12 layers, 12 attention heads, and an embedding dimension of 768. Modality-specific convolutional patch embedding layers use a patch size of $32\times 32$ for both images with input resolution $224\times 224$ and spectrograms with input resolution $256 \times 128$, where 256 corresponds to the time axis and 128 to the frequency axis. For the text modality, the sequences are truncated or padded to a fixed length of 256 tokens.

To mitigate potential modality imbalance during pretraining, we subsample the AudioSet and Wikipedia corpora to approximately 1.28 million samples for each that match the training size of ImageNet-1K. This balanced sampling ensures that each modality contributes equally to gradient updates across training iterations.

During pretraining, we apply modality-specific data augmentations. For images, random resized cropping, horizontal flipping, color jittering, and Gaussian blur are used base on the configuration in SimCLR \cite{chen2020simple}. For audio, time and frequency masking for spectrograms are used based on the configuration proposed in in \cite{rehman2025fssuavl, gong2022ssast}. For text, random token masking is used based on the setting in \cite{gao2021simcse}. We optimize Omni-C using AdamW \cite{loshchilov2017decoupled} with learning rate $1e-4$ and decay rate 0.1. The batch size for each modality is set 256.  We adopt a cosine learning rate schedule with warmup, where the minimum learning rate and epochs of warm up was set to $1e-5$ and 5, respectively. 

\noindent\textbf{Contrastive Loss.} We employ a contrastive loss function inspired by SimCLR \cite{chen2020simple} to learn the robust representations with each modality. For a given minibatch of size $N$ sampled from a single modality $m$, we generte two augmented views for each sample as described earlier. This process results in $2N$ projected embeddings. Let $z_{i}^{m}$ and $z_{j}^{m}$ denote the projected embeddings obtained via the modality-specific MLPs head. The contrastive loss for the positive pair $(i,j)$ in modality $m$ is defined as
\begin{equation}
\ell_{i,j}^{m} = -\log \frac{\exp(\text{sim}(\mathbf{z}_{i}^{m}, \mathbf{z}_{j}^{m})/\tau)}{\sum_{k=1}^{2N} \exp(\text{sim}(\mathbf{z}_{i}^{m}, \mathbf{z}_{k}^{m})/\tau)}
\label{eq:contrastive_loss}
\end{equation}
where the temperature parameter $\tau$ set to 0.05 and the sum in the denominator includes all $2N$ embeddings in the minibatch except the identity term for $k=i$. The cosine similarity is given by $sim(u, v)= u^{T}v/(||u|| ||v||)$
where $u$ and $v$ are the projected embeddings of two different views. This objective encourages similarity between augmented views of the same instance while pushing apart representations of different instances, fostering invariant and discriminative features.

\noindent\textbf{Pretrained model evaluation details.} We evaluate the Omni-C pretrained model on a range of downstream classification tasks. For this purpose, we use multiple protocols. For the unimodel classification tasks, we perform the evaluation using, (1)zero-shot (2) Linear probe, and (3) SBoRA PEFT \cite{po2024sbora}. For the cross cross-modality alignment, we use SAIL \cite{zhang2025assessing} and evaluate the corresponding performance on zero-shot classification. 

For all downstream evaluation tasks except zero-shot evaluation, we finetune Omni-C using AdamW with a learning rate of 1e-4, weight decay of 0.1, a cosine learning rate schedule with 10 warmup epochs and a minimum learning rate of 1e-5, and a total of 40 training epochs. 

For linear probe, we follow the typical protocol, as adopted in \cite{chen2020simple}, \cite{rehman2025fssuavl}, \cite{xie2022should}, \cite{rehman2024exploring}, that involves removing the projection head from the pretrained backbone and initializing a new linear classification layer, which is trained on top of the frozen class token CLS representations. 

For fine-tuning, we adopt the SBoRA method \cite{po2024sbora}, a parameter-efficient adaptation technique that extends LoRA \cite{hu2022lora}. By leveraging orthogonal standard basis vectors to initialize one of the low-rank matrices, SBoRA enables regional (sparse) weight updates and activates only a small fraction of the backbone parameters—approximately 12\% in our configuration—thereby reducing the risk of overfitting on downstream tasks. Specifically, we employ the SBoRA-FA variant with rank and scaling factor alpha set to 128. The scaling factor controls the magnitude of the low-rank updates added to the frozen pretrained weights. Following the recommendations in the original SBoRA \cite{po2024sbora}, we apply zero dropout immediately before the SBoRA layers.

\begin{table*}[t]
    \caption{Evaluation for \textbf{zero shot on image downstream task} for contrastive pretrained ViT-Base-32 model. I, A, and T denote Image, Audio, and Text, respectively.}
    \centering
    
    \resizebox{2\columnwidth}{!}{
        \begin{tabular}{l c c c c c c c c c c}
        \hline
             Model & Cars & DTD & EuroSAT & GTSRB  & KITTI & MNIST & RESISC45 & SUN397 & SVHM & \textbf{Avg Acc} \\
            \hline

            \cellcolor{lightgray}Expert-Image & \cellcolor{lightgray}2.46 & \cellcolor{lightgray}\textbf{38.28} & \cellcolor{lightgray}72.86 & \cellcolor{lightgray}22.13 & \cellcolor{lightgray}\textbf{64.21} & \cellcolor{lightgray}39.68 & \cellcolor{lightgray}41.85 & \cellcolor{lightgray}\textbf{26.07} & \cellcolor{lightgray}20.10 & \cellcolor{lightgray}\textbf{36.40} \\
            
            Omni-C (I \& A) & 2.16 & 34.10 & 73.79 & 21.67 & 57.03 & 39.89 & 43.12 & 21.70 & \textbf{22.48} & 35.10 \\

            Omni-C (I \& T) & 2.34 & 34.26 & 73.89 & \textbf{22.22} & 54.69 & 47.36 & \textbf{43.53} & 21.47 & 20.33 & 35.56 \\
            
            Omni-C (I \& A \& T) & 2.22 & 33.37 & \textbf{74.97} & 21.40 & 57.18 & \textbf{47.84} & 42.93 & 20.01 & 21.80 & 35.74 \\
            \hline
            
        \end{tabular}
    }
    \label{tab:downstream-image-zs}
\end{table*}

\begin{table}[t]
    \caption{Evaluation for \textbf{zero shot on audio downstream task} for contrastive pretrained ViT-Base-32 model. I, A, and T denote Image, Audio, and Text, respectively.}
    \centering
    \resizebox{1\columnwidth}{!}{
        \begin{tabular}{l c c c c c}
            \hline
             Model & VGGSound & EPIC-Sound & SpeechCommand & Nsynth  & \textbf{Avg Acc} \\
            \hline

            \cellcolor{lightgray}Expert-Audio & \cellcolor{lightgray}\textbf{5.80} & \cellcolor{lightgray}9.17 & \cellcolor{lightgray}\textbf{9.48} & \cellcolor{lightgray}\textbf{27.53} & \cellcolor{lightgray}\textbf{12.99}  \\

            Omni-C (I \& A) & 2.89 & 4.64 & 8.81 & 20.63 & 9.24  \\

            Omni-C (A\& T) & 4.86 & \textbf{9.19} & 7.47 & 25.63 & 11.78 \\

            Omni-C (I \& A \&T) & 2.63 & 4.61 & 9.00 & 23.43 & 9.91 \\

            \hline
            
        \end{tabular}
    }
    \label{tab:downstream-audio-zs}
\end{table}

\begin{table}[t]
    \caption{Evaluation for \textbf{zero shot on text downstream task} for contrastive pretrained ViT-Base-32 model. I, A, and T denote Image, Audio, and Text, respectively.}
    \centering
    \resizebox{1\columnwidth}{!}{
        \begin{tabular}{l c c c c c c}
            \hline
            Model & AGNEWS & NEWSGROUPS20 & IMDB & CARER  & \textbf{Avg Acc} \\
            \hline
            
            \cellcolor{lightgray}Expert-Text & \cellcolor{lightgray}\textbf{80.75} & \cellcolor{lightgray}\textbf{16.30} & \cellcolor{lightgray}52.00 & \cellcolor{lightgray}\textbf{21.82} & \cellcolor{lightgray}\textbf{42.71}  \\
            
            Omni-C (I \& T) & 49.54 & 13.60 & \textbf{55.00} & 18.07 & 34.05  \\

            Omni-C (A \& T) & 53.30 & 12.06 & 54.14 & 20.78 & 35.07  \\

            Omni-C (I \& A \& T) & 56.08 & 12.82 & 52.85 & 17.08 & 34.70  \\
            \hline

        \end{tabular}
    }
    \label{tab:downstream-text-zs}
\end{table}

\begin{table}[t]
    \centering
    \caption{Downstream tasks datasets used to evaluate Omni-C on image, audio and text modalities. The table reports the task, number of classes (\# cls), number of training samples (\# train), and the number of validation samples (\# valid) for each datasets.}
    \resizebox{1\columnwidth}{!}{
    \begin{tabular}{|l|c|c|c|c|c}
        \hline
        \textbf{Dataset}  & \textbf{Tasks} & \textbf{\# cls} & \textbf{\# train} & \textbf{\# valid}  \\
        \hline
        Cars \cite{krause20133d} & Fine-grained car cls. & 196 & 6K & 1.7K  \\
        DTD \cite{cimpoi2014describing} & Texture cls. & 47 & 1.8K  & 1.8K  \\
        EuroSAT \cite{helber2019eurosat} & Land use and land cover cls. & 10 & 18K & 4K \\
        GTSRB \cite{haloi2015traffic} & Traffic Sign cls. & 43 & 31K & 7.8K \\
        KITTI \cite{geiger2012we} & Autonomous driving car cls. & 9 & 9.6K & 2K\\
        MNIST \cite{deng2012mnist} & Hand written digit cls. & 10 & 48K & 12K\\
        RESISC45 \cite{cheng2017remote} & Remote sensing image cls. & 45 & 13K & 2.8K \\
        SUN397 \cite{xiao2016sun} & Scene understanding cls. & 397 & 76K & 10.7K \\
        SVHN \cite{goodfellow2013multi} & House number cls. & 10 & 58K & 15K \\
        \hline
        VGGSound \cite{chen2020vggsound} & Audio Event cls. & 309 & 183K & 15K \\
        EPIC-Sound \cite{huh2025epic} & Egocentric sound event cls. & 44 & 60K & 8K \\
        SpeechCommand \cite{warden2018speech} & Keyword spotting cls. & 35 & 166K & 10K \\
        Nsynth \cite{engel2017neural} & Musical instrument cls. & 11 & 289K & 12K \\
        \hline
        AGNews \cite{yogatama2017generative} & News Topic cls. & 4 & 120K & 7.6K \\
        Newsgroups20 \cite{albishre2015effective} & Document categorization & 20 & 11K & 7.5K \\
        IMDB \cite{tripathi2020analyzing}  & Sentiment analysis & 2 & 40K & 10K \\
        CARER \cite{saravia2018carer}  & Emotion cls. & 6 & 16K & 2K \\
        \hline
        \end{tabular}
    }
    \label{tab:downstream-datasets}
\end{table}

\noindent\textbf{Evaluation datasets.} We evaluate Omni-C (pretrained on Image \& Audio \& Text) model on a diverse set of image, audio, and text downstream tasks. The summary of the datasets for the downstream tasks are listed in Table \ref{tab:downstream-datasets}. The details will be provided in the following subsection.

\noindent\textbf{Images}. Evaluation of Omni-C on image-based downstream tasks inlcude fine-grained vechicle recognition (Standard Cars dataset \cite{krause20133d}), texture classification (Describable Textures Dataset (DTD) \cite{cimpoi2014describing}), land use and land cover classification (EuroSAT \cite{helber2019eurosat}), traffic sign recognition ( German Traffic Sign Recognition Benchmark (GTSRB) \cite{haloi2015traffic}), object classification in autonomous driving scenes (KITTI \cite{geiger2012we}), handwritten digit classification (MNIST \cite{deng2012mnist}), remote sensing scene classification (RESISC45 \cite{cheng2017remote}), scene recognition like airport terminal, tower, and etc (SUN397 \cite{xiao2016sun}), and house number digit classification (SVHN \cite{goodfellow2013multi}).

\noindent\textbf{Audio.} We evaluate Omni-C on the VGGSound dataset \cite{chen2020vggsound}, which emphasizes audio-visual event recognition; the EpicSounds dataset \cite{huh2025epic} that focuses on egocentric action sounds; the Speech Commands dataset \cite{warden2018speech} for keyword spotting; and the NSynth dataset \cite{engel2017neural} for musical instrument note classification.

\noindent\textbf{Text.} For text-based downstream tasks, Omni-C is evalauted on news topic classification(AGNews \cite{yogatama2017generative}), document categorization (Newsgroups20 \cite{albishre2015effective}), sentiment analysis (IMDB \cite{tripathi2020analyzing}), and clinical assertion status detection (CARER \cite{saravia2018carer}).

\noindent\textbf{Evaluation Metrices.} For all unimodal downstream classification tasks, we report top-1 accuracy on the standard test or validation splits. For cross-modal evaluation, we report the CLIP's style zero-shot performance on image-text and audio-text downstream classification tasks after the SAIL \cite{zhang2025assessing} alignment.

\begin{figure*}[ht]
      \centering
      \subfloat[Expert Image]{\includegraphics[width=0.3\linewidth, height=0.25\linewidth]{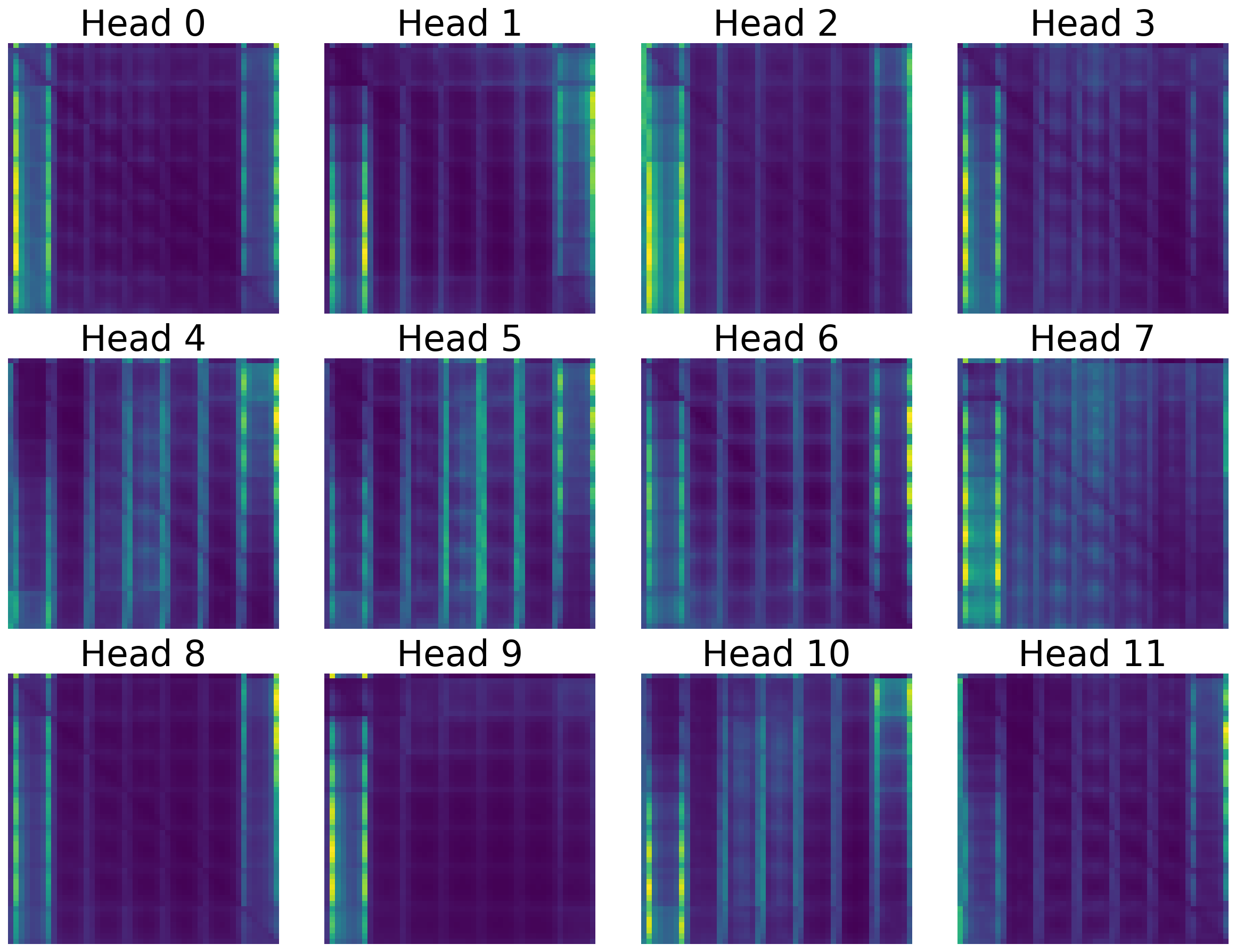}
      \label{fig:attn-map-expert-image}}
      \subfloat[Expert Audio]{\includegraphics[width=0.3\linewidth,height=0.25\linewidth]{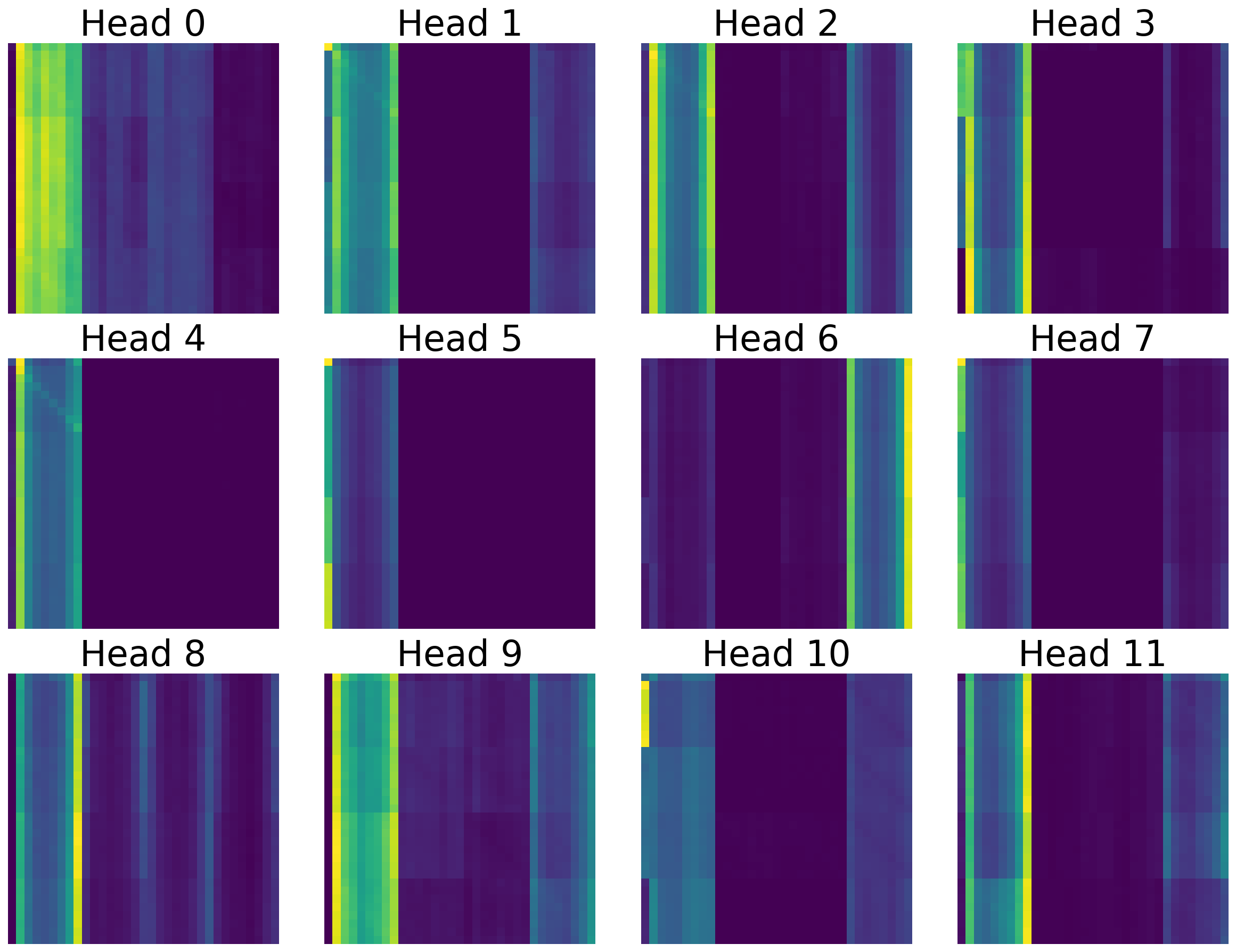}
        \label{fig:attn-map-expert-audio}}
      \subfloat[Expert Text]{\includegraphics[width=0.3\linewidth,height=0.25\linewidth]{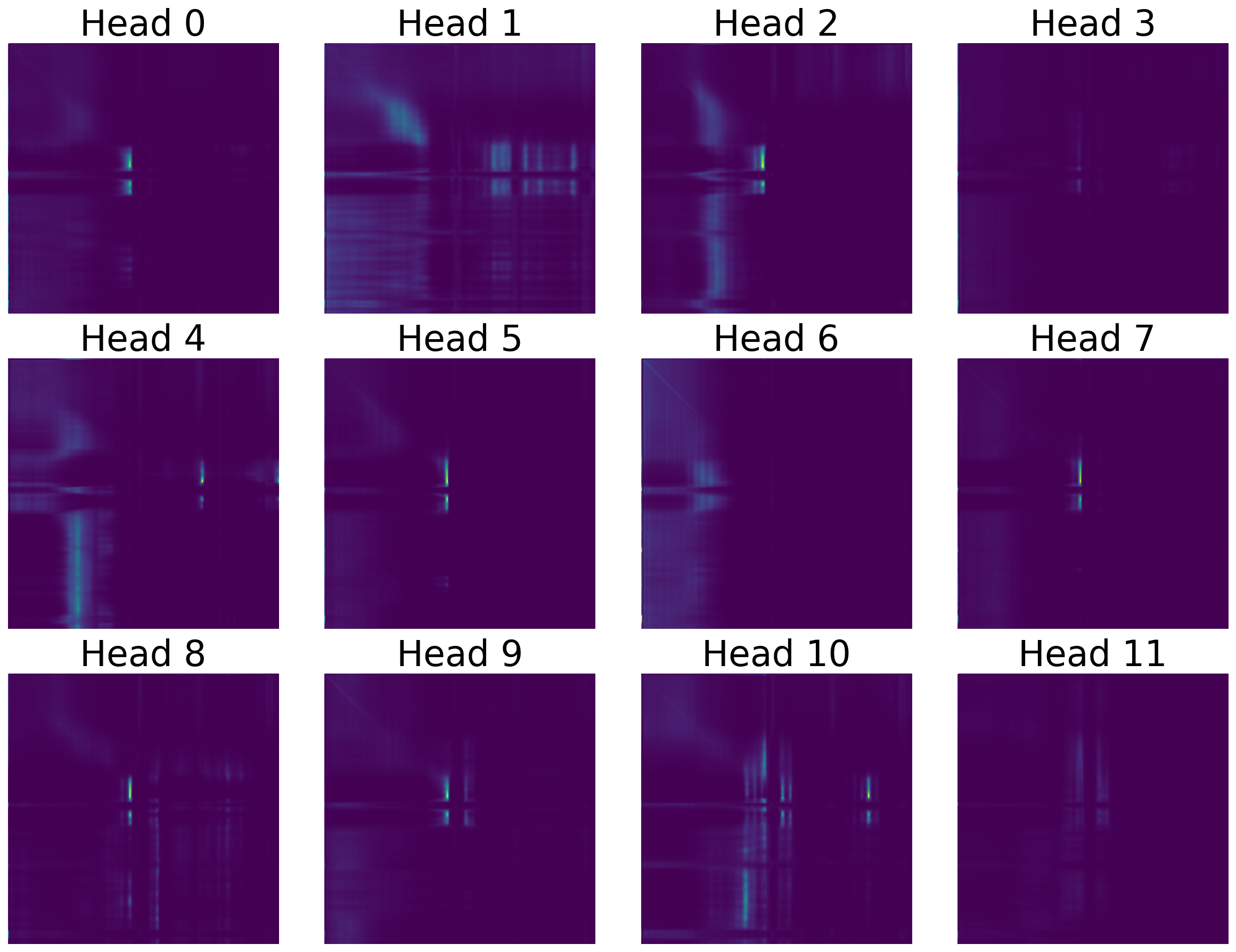}
        \label{fig:attn-map-expert-text}}\\
      \subfloat[Omni-C (Image)]{\includegraphics[width=0.3\linewidth,height=0.25\linewidth]{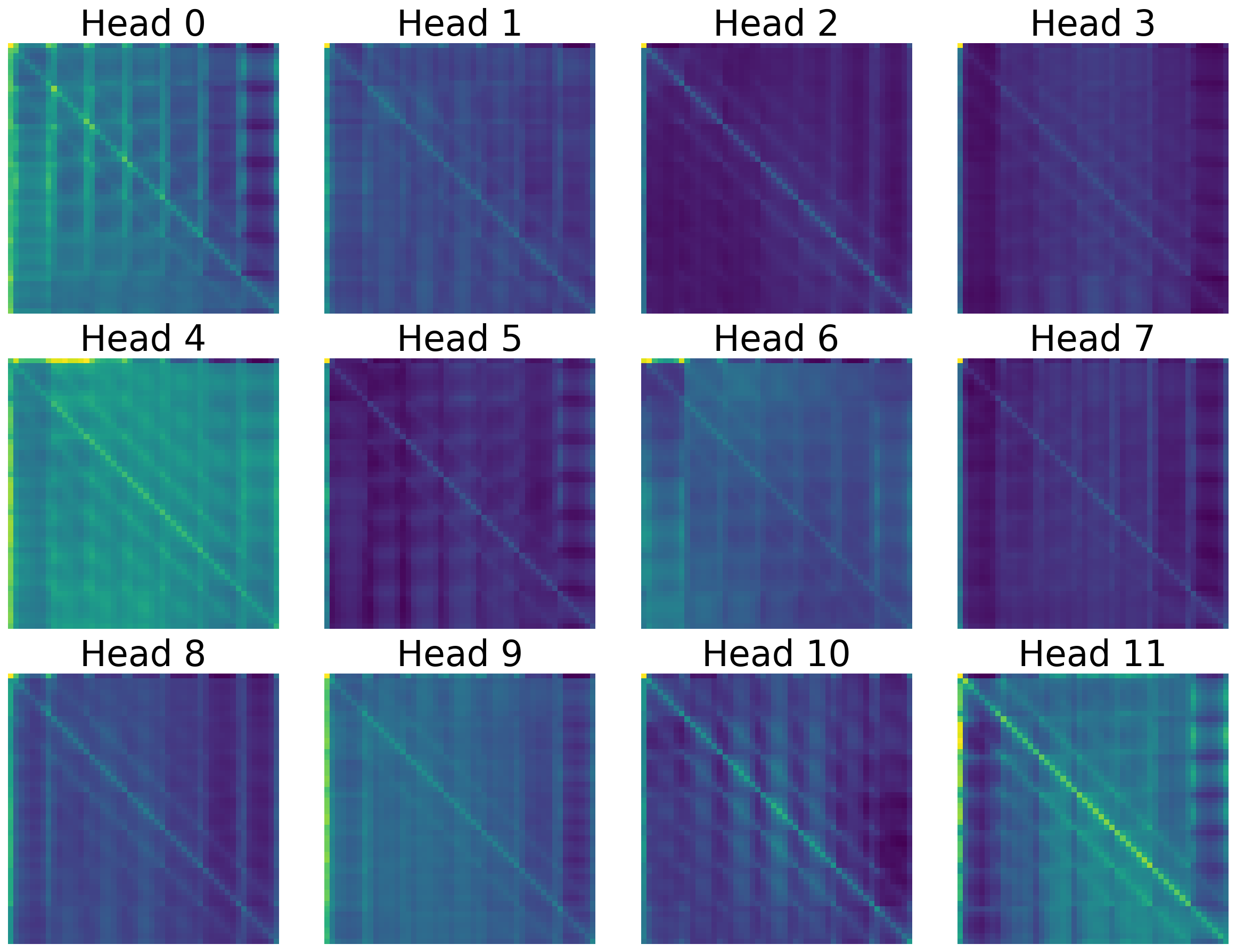}
          \label{fig:attn-map-omni-image}
        }
        \subfloat[Omni-C (Audio)]{\includegraphics[width=0.3\linewidth,height=0.25\linewidth]{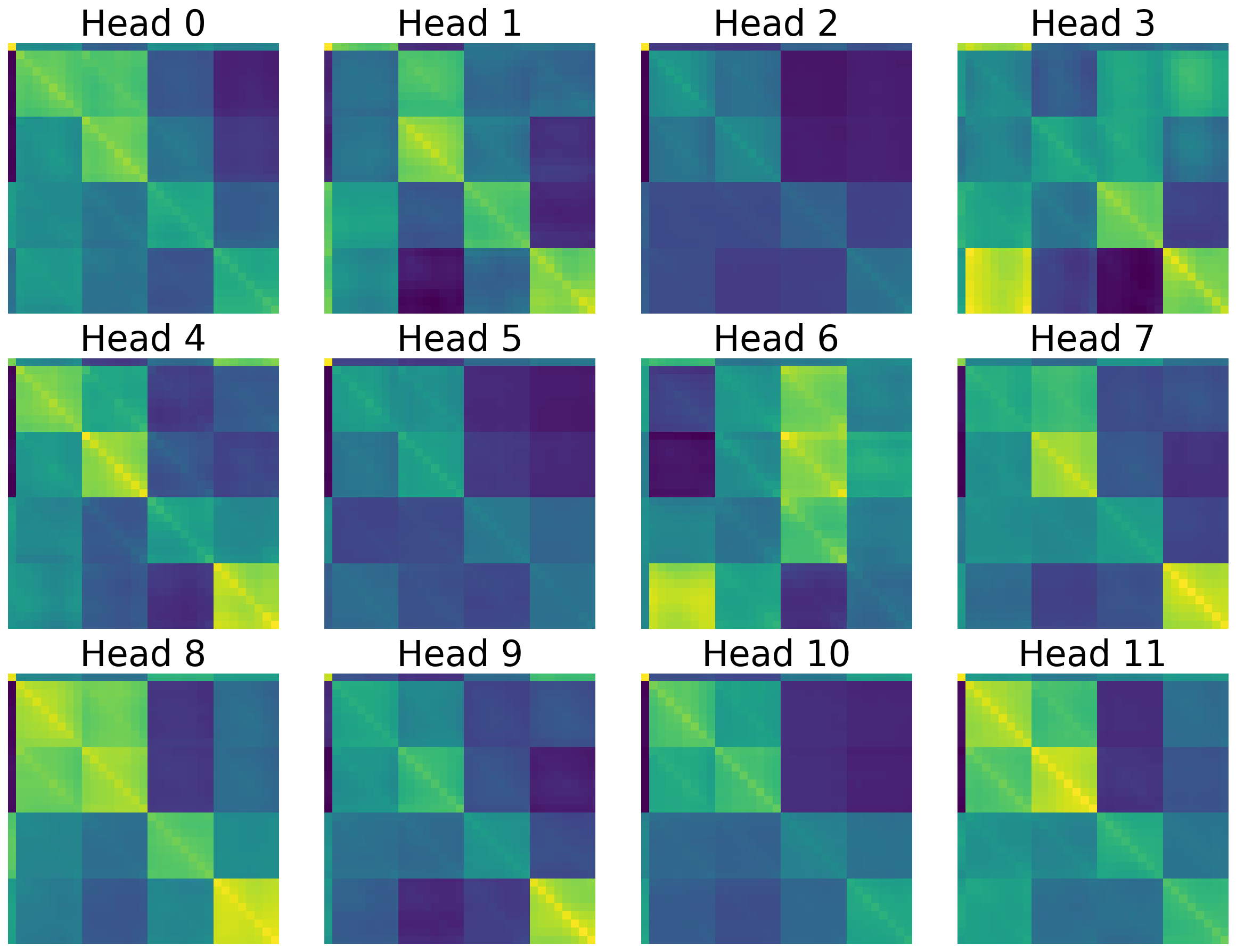}
              \label{fig:attn-map-omni-audio}
            }
        \subfloat[Omni-C (Text)]{\includegraphics[width=0.3\linewidth,height=0.25\linewidth]{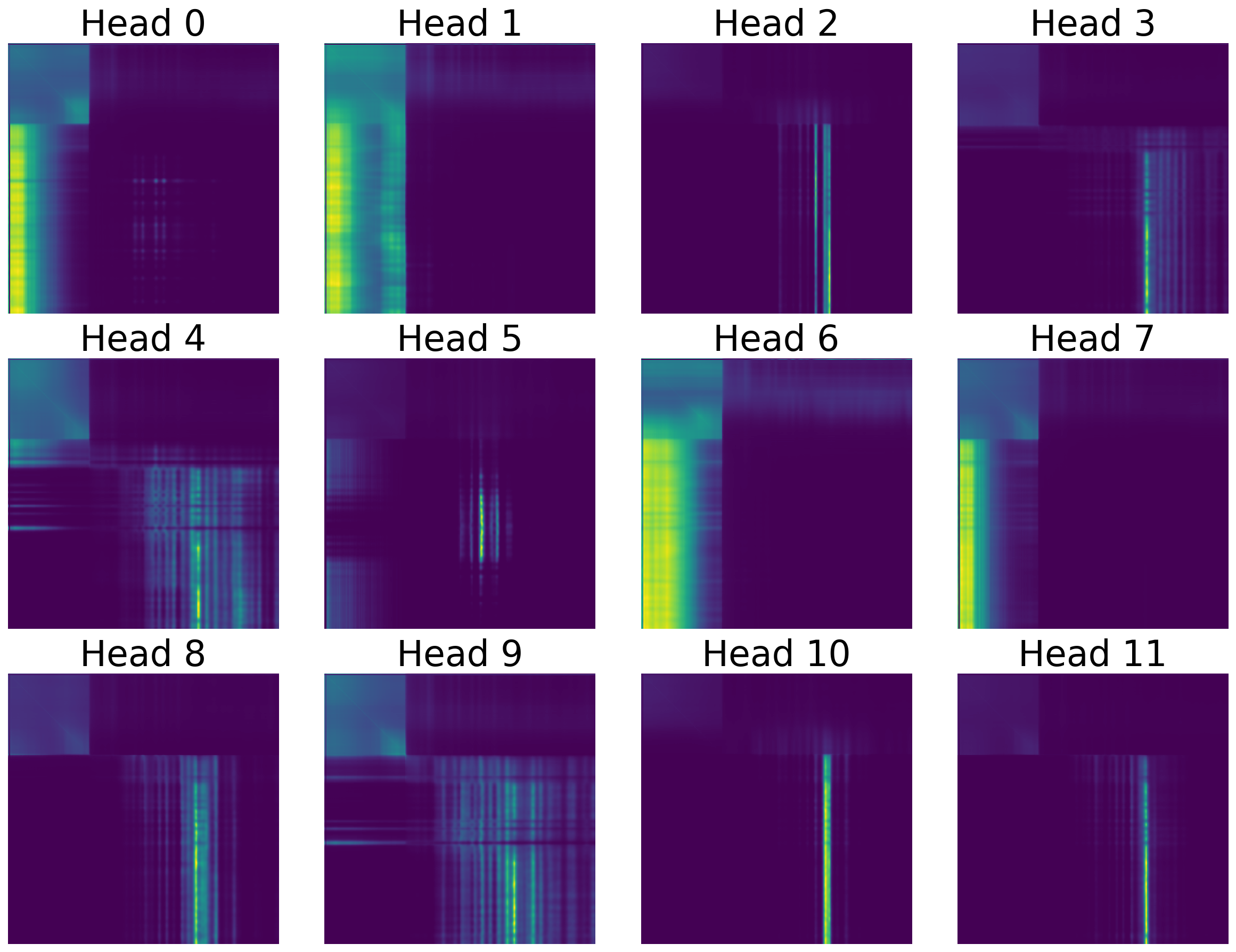}
              \label{fig:attn-map-omni-text}
            }
      \caption{Average self-attention maps from the last ViT-Base Transformer layer with 12 heads over 3000 samples \textbf{for the pretrained models} from downstream datasets. (a-c) show attention maps for the modality-specific expert models on images (KITTI), audio spectrograms (VGGSound), and text (AGNews), respectively, exhibiting focused attention patterns that specialize in modality-specific local features. (d-f) show corresponding attention maps for the unified Omni-C model on the same inputs and datasets, revealing distributed attention that concurrently encodes and integrates information from heterogeneous inputs.}
      \label{fig:attn-map-omni}
\end{figure*}

\begin{figure*}[ht]
      \centering
      \subfloat[Expert Image]{\includegraphics[width=0.3\linewidth, height=0.25\linewidth]{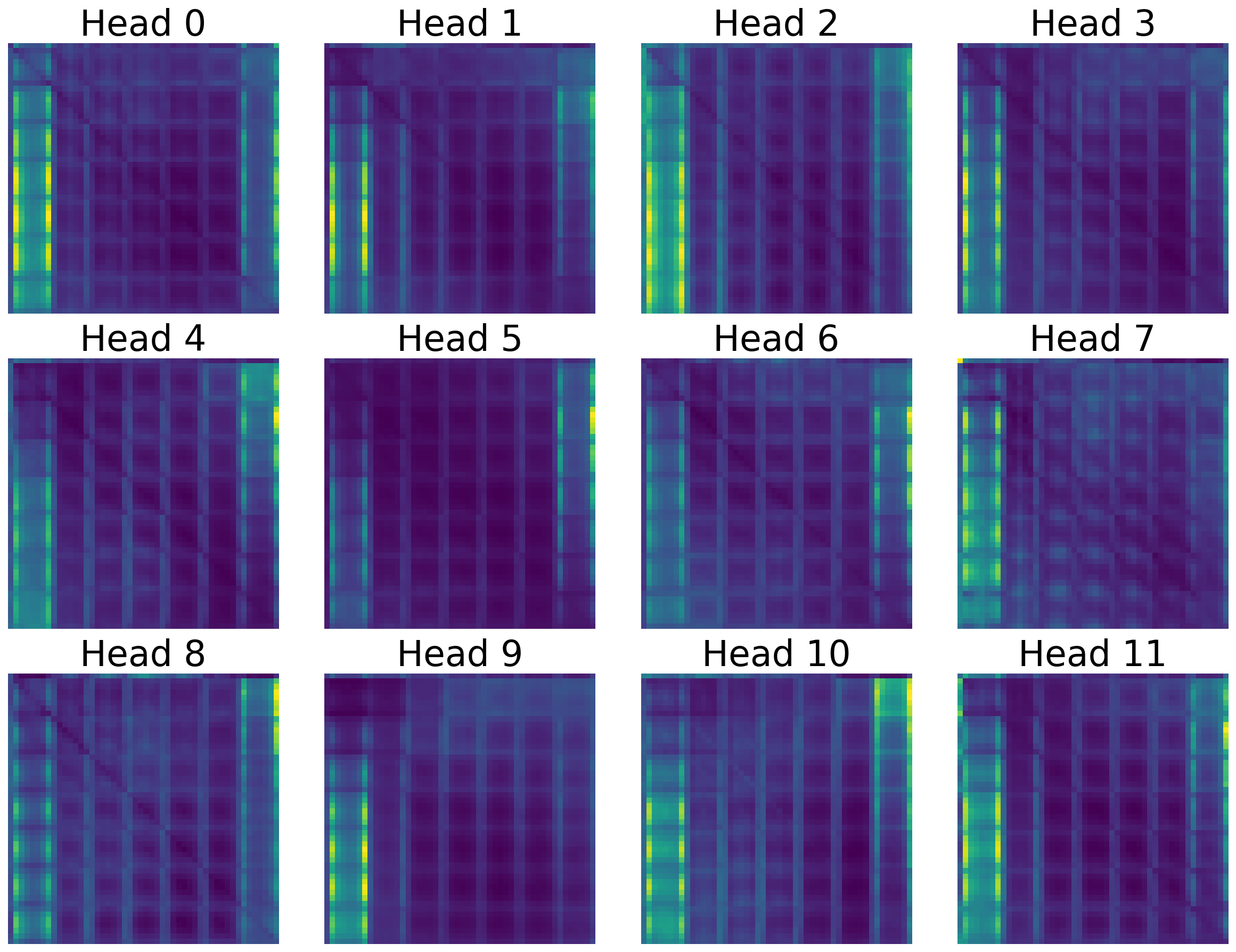}
      \label{fig:attn-map-expert-image-sbora}}
      \subfloat[Expert Audio]{\includegraphics[width=0.3\linewidth,height=0.25\linewidth]{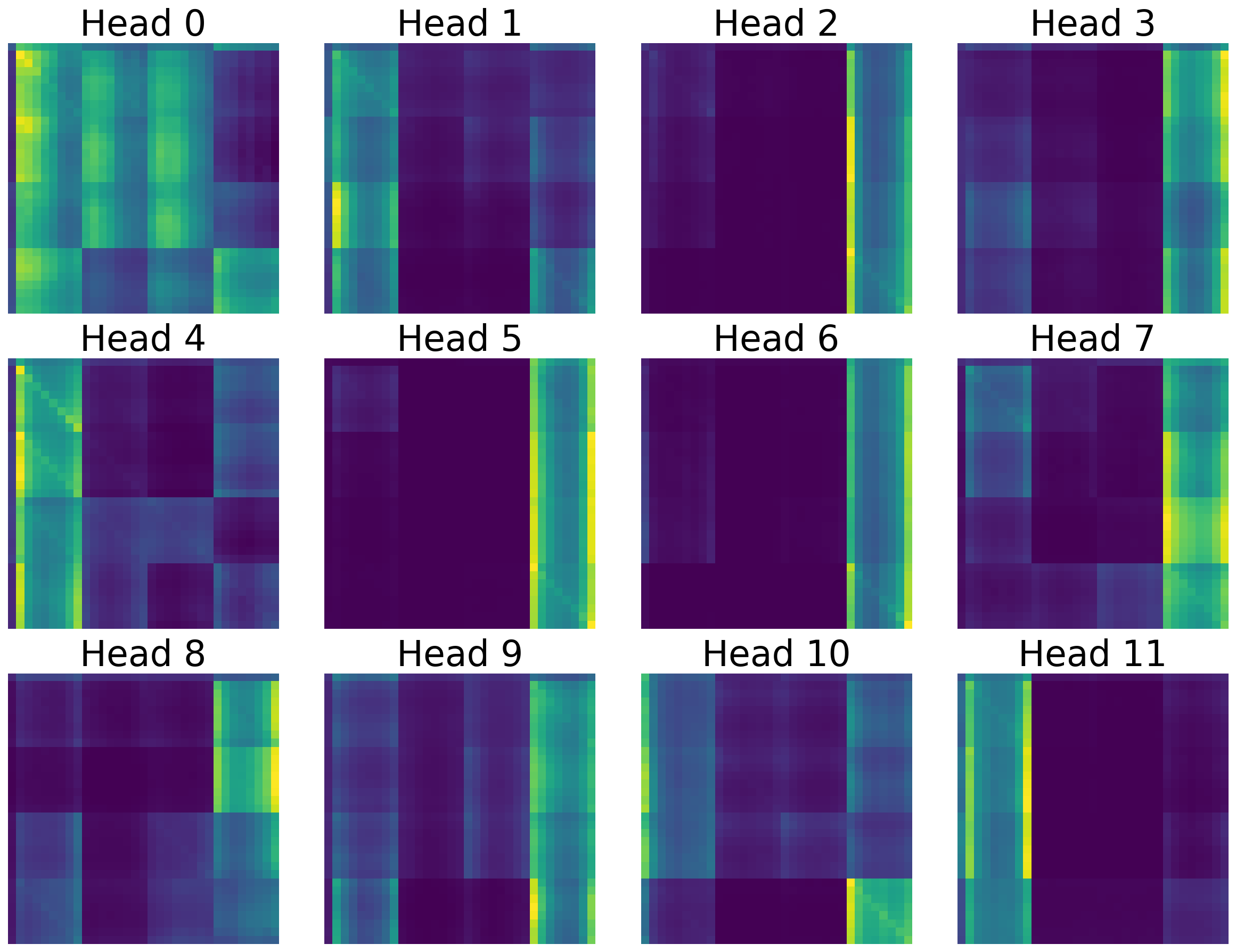}
        \label{fig:attn-map-expert-audio-sbora}}
      \subfloat[Expert Text]{\includegraphics[width=0.3\linewidth,height=0.25\linewidth]{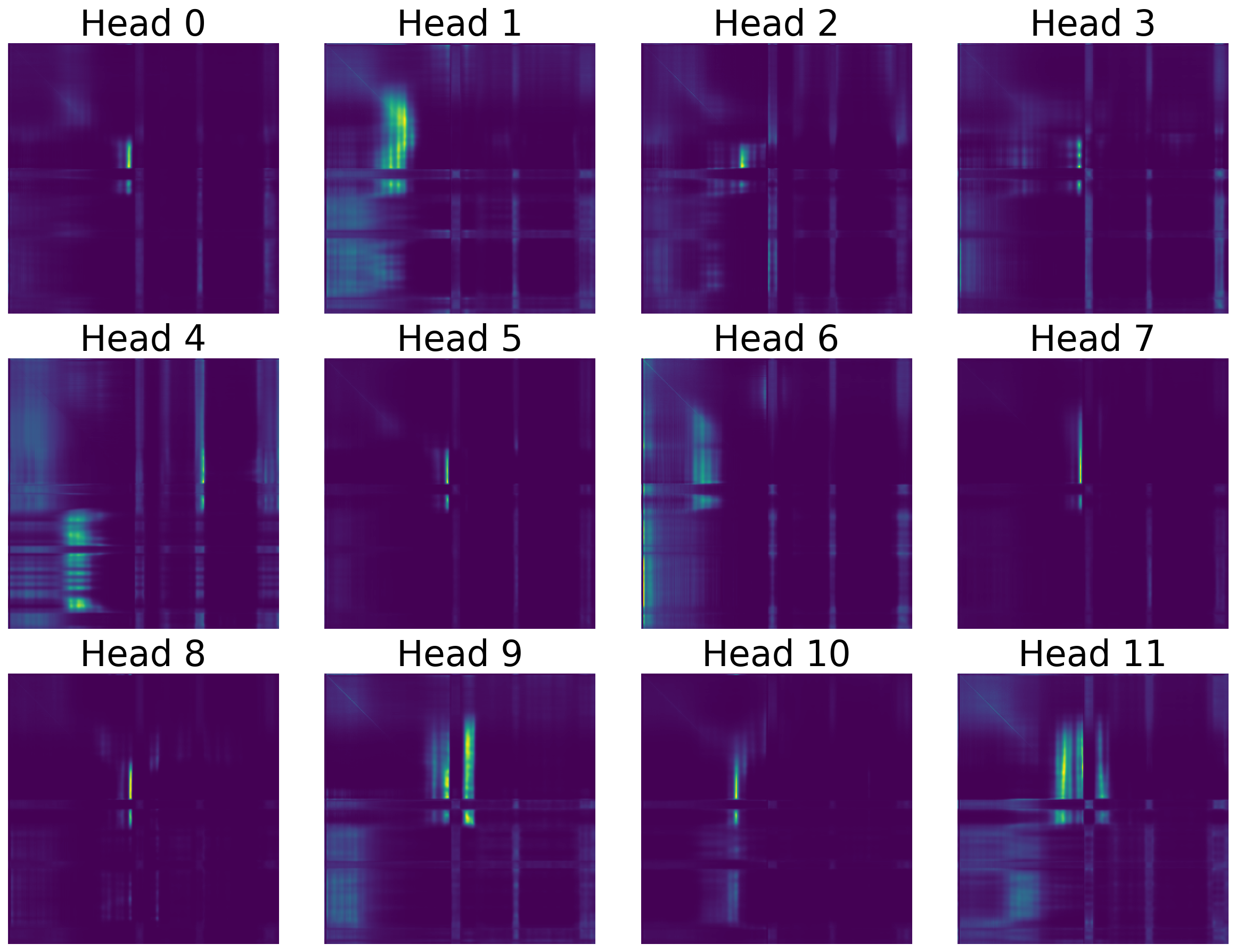}
        \label{fig:attn-map-expert-text-sbora}}\\
      \subfloat[Omni-C (Image)]{\includegraphics[width=0.3\linewidth,height=0.25\linewidth]{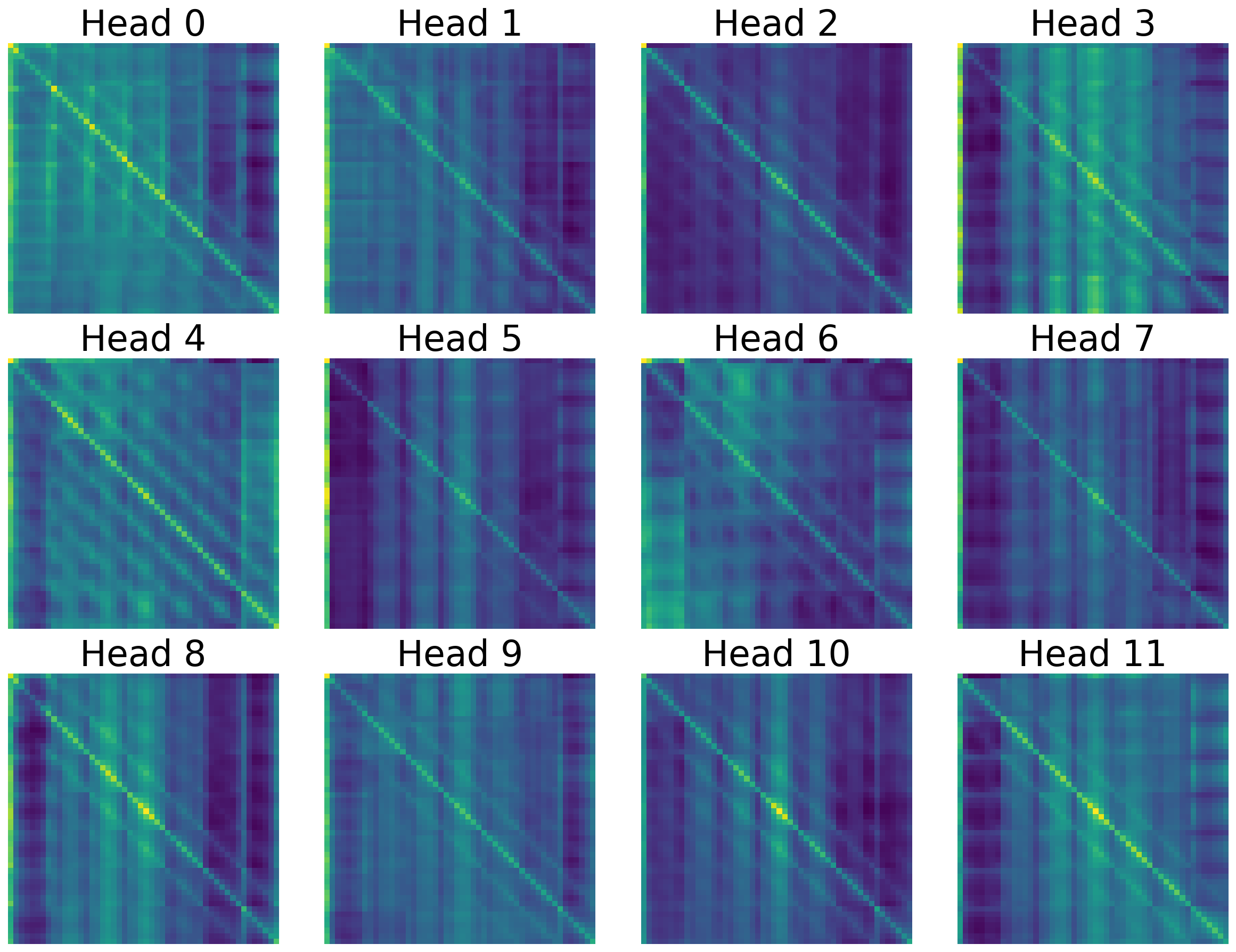}
          \label{fig:attn-map-omni-image-sbora}
        }
        \subfloat[Omni-C (Audio)]{\includegraphics[width=0.3\linewidth,height=0.25\linewidth]{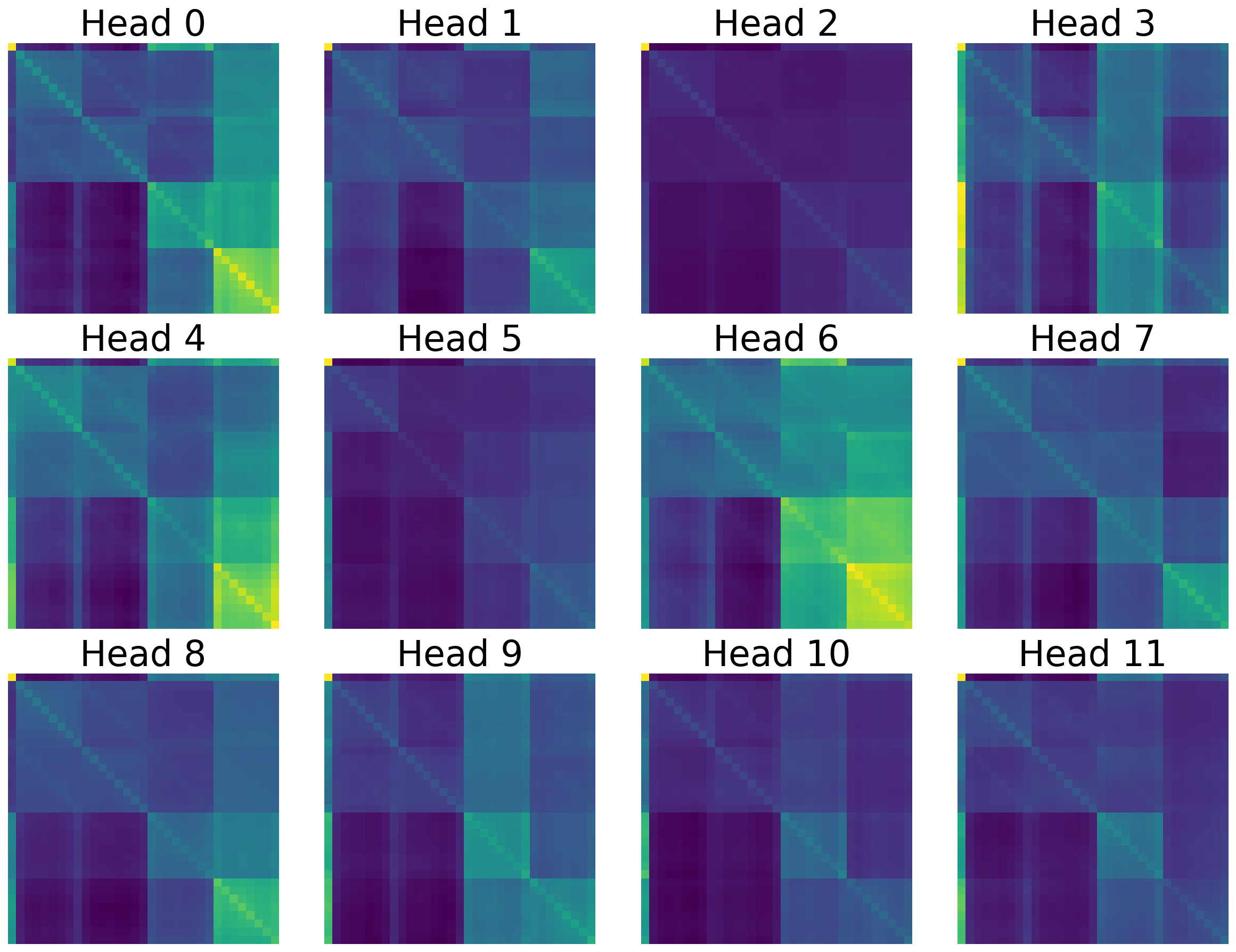}
              \label{fig:attn-map-omni-audio-sbora}
            }
        \subfloat[Omni-C (Text)]{\includegraphics[width=0.3\linewidth,height=0.25\linewidth]{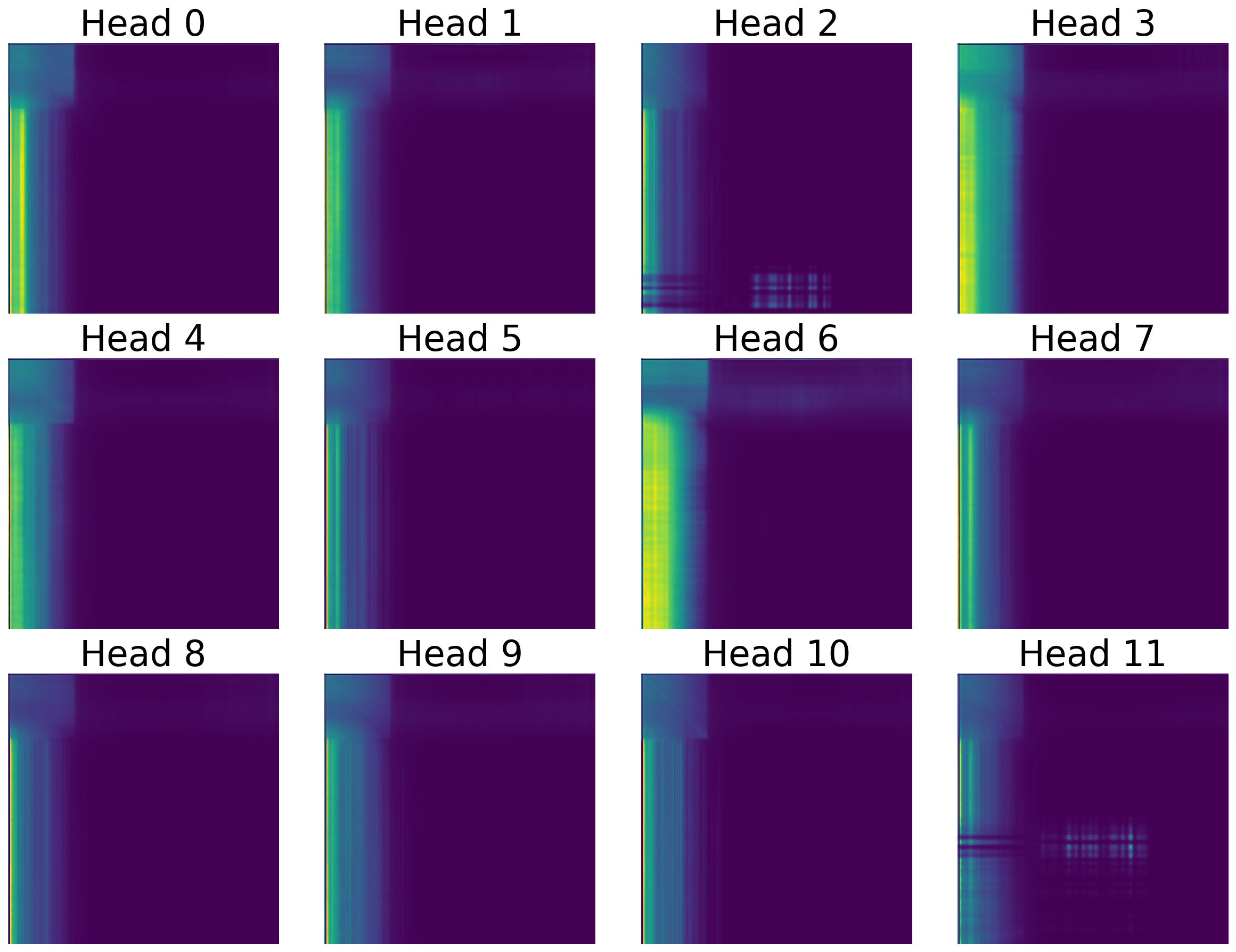}
              \label{fig:attn-map-omni-text-sbora}
            }
      \caption{Average self-attention maps from the last ViT-Base Transformer layer with 12 heads over 3000 samples \textbf{after SBoRA downstream datasets fine tuning}. (a-c) show attention maps for the modality-specific expert models on images (KITTI), audio spectrograms (VGGSound), and text (AGNews), respectively. (d-f) show corresponding attention maps for the unified Omni-C model on the same inputs and datasets. Importantly, the Omni-C backbone can effectively recover from its distributed attention (optimized for cross-modal generalization) to focused, modality-specific attention patterns through lightweight parameter-efficient fine-tuning (SBoRA)}
      \label{fig:attn-map-omni-sbora}
\end{figure*}

\begin{table*}[t]
    \caption{Evaluation for \textbf{linear probe on image downstream task} for contrastive pretrained ViT-Base-32 model. I, A, and T denote Image, Audio, and Text, respectively.}
    \centering
    
    \resizebox{2\columnwidth}{!}{
        \begin{tabular}{l c c c c c c c c c c}
        \hline
            Model & Cars & DTD & EuroSAT & GTSRB  & KITTI & MNIST & RESISC45 & SUN397 & SVHM & \textbf{Avg Acc} \\
            \hline
        
            \cellcolor{lightgray}Expert-Image & \cellcolor{lightgray}6.87 & \cellcolor{lightgray}\textbf{45.32} & \cellcolor{lightgray}\textbf{92.36} & \cellcolor{lightgray}76.76 & \cellcolor{lightgray}\textbf{92.24} & \cellcolor{lightgray}90.23 & \cellcolor{lightgray}\textbf{73.67} & \cellcolor{lightgray}\textbf{55.34} & \cellcolor{lightgray}\textbf{49.44} & \cellcolor{lightgray}\textbf{64.69} \\

            Omni-C (I \& A) & 9.35 & 43.53 & 91.36 & 75.15 & \textbf{92.24} & 91.96 & 71.98 & 51.22 & 47.34 & 63.79 \\

            Omni-C (I \& T) & 9.04 & 41.99 & 91.48 & \textbf{77.91} & 92.15 & 92.05 & 73.27 & 51.04 & 47.64 & 64.06 \\

            Omni-C (I \& A \& T) & 8.75 & 41.18 & 91.28 & 76.40 & 91.41 & \textbf{92.87} & 72.07 & 49.03 & 49.19 & 63.57 \\
            \hline
            
        \end{tabular}
    }
    \label{tab:downstream-image-lp}
\end{table*}

\subsection{Comparison with Modality-Specific Pretrain Model on Downstream Tasks}
\label{subsec:knn-zero-shot}
\noindent\textbf{Zero-shot evaluation.} First, we compare the Omni-C model against modality-specific baselines (experts) that are pretrained with unimodal contrastive loss on their respective datasets. One can see from Table \ref{tab:downstream-image-zs}, \ref{tab:downstream-audio-zs}, and \ref{tab:downstream-text-zs} that the Omni-C model obtained an average top-1 accuracy of 35.74\% on images (vs. 36.40\% for the image expert), 9.91\% on audio (vs. 12.99\% for the audio expert), and 34.70\% on text (vs. 42.71\% for the text expert). These results demonstrate near parity in zero-shot image performance and more pronounced degradation for audio and text modalities. We found these results encouraging due to the fact that the Omni-C model enforces distributed-attention in the input patches, enabling it encode information for multiple modalities at the same time than their unimodal counter parts that encode modality-specific information due to focus-attention. 

While the literature in ViTs overlooks distributed attention, we found numerous examples of such phenomena in perceptual psychology that distinguishes between distributed-attention and foucs-attention \cite{10.1093/acprof:oso/9780199658442.003.0002}. 
Distributed-attention spreads broadly across a scene to extract the global gist-like summaries, while focused attention narrows to individual elements for accurate identification. The shared Omni-C backbone enforces a distributed attention mode, in which attention spreads broadly across different patch token representations. This can be readily seen Figs.\ref{fig:attn-map-omni-image}, \ref{fig:attn-map-omni-audio}, and \ref{fig:attn-map-omni-text}. This broad distribution naturally aligns with the expert image model’s attention pattern very closely (Fig. \ref{fig:attn-map-expert-image}) and allows Omni-C to maintain robust global context information for achieving near-parity zero-shot image performance.

In contrast, expert audio and text models rely on more focused and specialized attention distributions (elongated stripes for audio’s spectro-temporal structure in figure \ref{fig:attn-map-expert-audio} and sparse vertical patterns for text’s sequential dependencies in figure \ref{fig:attn-map-expert-text}). The distributed attention properties induced by cross-modal training distorts these specialized patterns in Omni-C (figure \ref{fig:attn-map-omni-audio} and figure \ref{fig:attn-map-omni-text}) and leads to more performance degradation on audio (around 3\% drop) and text tasks (around 8\% drop). However, these modality-specific local details captured from the focus attentions can be largely recovered by linear probing and fine-tuning the Omni-C model.

\noindent\textbf{Linear probe evaluation.} In linear probe (Tables \ref{tab:downstream-image-lp}, \ref{tab:downstream-audio-lp}, and \ref{tab:downstream-text-lp}), the Omni-C model achieves a comparable performance or outperform to the modality-specific experts across all modalities. The unified model attains average top-1 accuracies of 63.57\% for images (versus 64.69\% for the expert), 34.85\% for audio (versus 33.12\% for expert), and 61.87\% for text (versus 61.65\% for expert). These results stand in contrast to the modest zero-shot degradations observed, particularly in audio and text. This near-equivalent performance under linear probe, where the pretrained backbone is frozen and only a simple linear classifier is trained on the global CLS token representations, demonstrates that the Omni-C model effectively captures generic, transferable features through its pretraining. This further shows that distributed-attention can indeed help in learning multiple heterogeneous modalities using the same architecture. The internal neural pathways, induced by dense parameter sharing, retain high-level information sufficient for strong downstream transfer when minimal adaptation is applied, supporting our claim that a single shared encoder can serve as an efficient universal approximator without substantial performance compromise.

\begin{table}[!t]
    \caption{Evaluation for \textbf{linear probe on audio downstream task} for contrastive pretrained ViT-Base-32 model. I, A, and T denote Image, Audio, and Text, respectively.}
    \centering
    \resizebox{1\columnwidth}{!}{
        \begin{tabular}{l c c c c c}
            \hline
             Model & VGGSound & EPIC-Sound & SpeechCommand & Nsynth  & \textbf{Avg Acc} \\
            \hline

            \cellcolor{lightgray}Expert-Audio & \cellcolor{lightgray}15.62 & \cellcolor{lightgray}32.87 & \cellcolor{lightgray}35.15 & \cellcolor{lightgray}48.84 & \cellcolor{lightgray}33.12  \\
            
            Omni-C (I \& A) & 18.13 & \textbf{34.04} & 36.61 & \textbf{55.89} & 36.16 \\
            
            Omni-C (A \& T) & 15.57 & 32.10 & \textbf{39.28} & 50.90 & 34.46 \\

            Omni-C (I \& A \& T) & \textbf{17.12} & 32.68 & 34.98 & 54.64 & \textbf{34.85}  \\
            \hline

        \end{tabular}
    }
    \label{tab:downstream-audio-lp}
\end{table}

\begin{table}[t]
    \caption{Evaluation for \textbf{linear probe on text downstream task} for contrastive pretrained ViT-Base-32 model. I, A, and T denote Image, Audio, and Text, respectively.}
    \centering
    \resizebox{1\columnwidth}{!}{
        \begin{tabular}{l c c c c c}
            \hline
             Model & AGNEWS & NEWSGROUPS20 & IMDB & CARER  & \textbf{Avg Acc} \\
            \hline

            \cellcolor{lightgray}Expert-Text &  \cellcolor{lightgray}\textbf{88.81} & \cellcolor{lightgray}45.03 & \cellcolor{lightgray}67.35 & \cellcolor{lightgray}42.40 & \cellcolor{lightgray}60.89  \\
            
            Omni-C (I \& T) & 87.95 & \textbf{46.32} & 67.89 & 45.25 & \textbf{61.85}  \\

            Omni-C (A \& T) & 87.76 & 44.78 & \textbf{69.16} & \textbf{45.71} & \textbf{61.85}  \\
            
            Omni-C (I \& A \& T) & 88.22 & 45.36 & 68.13 & 43.68 & 61.34  \\
            \hline
            
        \end{tabular}
    }
    \label{tab:downstream-text-lp}
\end{table}

\begin{table}[!ht]
    \caption{Evaluation for \textbf{SBoRA fine-tuning on audio downstream task} for contrastive pretrained ViT-Base-32 model. I, A, and T denote Image, Audio, and Text, respectively.}
    \centering
    \resizebox{1\columnwidth}{!}{
        \begin{tabular}{l c c c c c}
            \hline
             Model & VGGSound & EPIC-Sound & SpeechCommand & Nsynth  & \textbf{Avg Acc} \\
            \hline

            \cellcolor{lightgray}Expert-Audio & \cellcolor{lightgray}\textbf{38.28} & \cellcolor{lightgray}\textbf{44.92} & \cellcolor{lightgray}89.04 & \cellcolor{lightgray}72.04 & \cellcolor{lightgray}\textbf{61.07}  \\

            Omni-C (I \& A) & 34.03 & 39.69 & 85.84 & 72.17 & 57.93  \\

            Omni-C (A \& T) & 33.82 & 40.21 & \textbf{89.60} & 72.08 & 58.92  \\
            
            Omni-C (I \& A \& T) & 33.39 & 39.07 & 87.61 & \textbf{72.47} & 58.13  \\
            \hline
            
        \end{tabular}
    }
    \label{tab:downstream-audio-sbora}
\end{table}

\begin{table}[!ht]
    \caption{Evaluation for \textbf{SBORA fine-tuning on text downstream task} for contrastive pretrained ViT-Base-32 model. I, A, and T denote Image, Audio, and Text, respectively.}
    \centering
    \resizebox{1\columnwidth}{!}{
        \begin{tabular}{l c c c c c}
            \hline
            Model & AGNEWS & NEWSGROUPS20 & IMDB & CARER  & \textbf{Avg Acc} \\
            \hline

            \cellcolor{lightgray}Expert-Text & \cellcolor{lightgray}\textbf{94.30} & \cellcolor{lightgray}\textbf{65.62} & \cellcolor{lightgray}\textbf{85.27} & \cellcolor{lightgray}\textbf{92.02} & \cellcolor{lightgray}\textbf{84.30}  \\

            Omni-C (I \& T) & 93.66 & 62.65 & 82.12 & 89.75 & 82.04  \\

            Omni-C (A \& T) & 94.00 & 61.50 & 81.94 & 89.67 & 81.77  \\

            Omni-C (I \& A \& T) & 93.72 & 60.25 & 82.73 & 90.24 & 81.73  \\
            \hline
            
        \end{tabular}
    }
    \label{tab:downstream-text-sbora}
\end{table}

\begin{table*}[t]
    \caption{Evaluation for \textbf{SBoRA fine-tuning on image downstream task} for contrastive pretrained ViT-Base-32 model. I, A, and T denote Image, Audio, and Text, respectively.}
    \centering
    
    \resizebox{2\columnwidth}{!}{
        \begin{tabular}{l c c c c c c c c c c}
        \hline
            Model & Cars & DTD & EuroSAT & GTSRB  & KITTI & MNIST & RESISC45 & SUN397 & SVHM & \textbf{Avg Acc} \\
            \hline
           
            \cellcolor{lightgray}Expert-Image & \cellcolor{lightgray}\textbf{53.57} & \cellcolor{lightgray}\textbf{57.42} & \cellcolor{lightgray}98.47 & \cellcolor{lightgray}99.95 & \cellcolor{lightgray}98.06 & \cellcolor{lightgray}99.51 & \cellcolor{lightgray}\textbf{90.18} & \cellcolor{lightgray}\textbf{64.90} & \cellcolor{lightgray}\textbf{95.03} & \cellcolor{lightgray}\textbf{84.12} \\

            Omni-C (I \& A) & 49.93 & 55.99 & 98.57 & 99.88 & \textbf{98.46} & 99.48 & 88.62 & 61.36 & 94.50 & 82.97 \\

            Omni-C (I \& T) & 47.81 & 55.99 & \textbf{98.69} & 99.83 & 98.27 & \textbf{99.59} & 88.97 & 60.69 & 94.41 & 82.69 \\

            Omni-C (I \& A \& T) & 45.99 & 53.36 & 98.22 & 99.88 & 98.41 & 99.54 & 88.30 & 60.31 & 94.60 & 82.06 \\
            \hline 
            
        \end{tabular}
    }
    \label{tab:downstream-image-sbora}
\end{table*}

\noindent\textbf{SBoRA efficient fine tuning.} Similarly, under SBoRA fine-tuning \cite{po2024sbora} (Tables \ref{tab:downstream-audio-sbora},  \ref{tab:downstream-text-sbora} and \ref{tab:downstream-image-sbora}), Omni-C achieves similar accuracy with the experts: 82.06\% on images (vs. 84.12\% for image expert), 58.13\% on audio (vs. 61.07\% for audio expert), and 81.79\% on text (vs. 84.30\% for text expert). These outcomes further highlight the model's ability to provide global and refinable representations, as the low-rank adaptation of SBoRA—activating only approximately 12\% of the backbone parameters. This process recovers any modality-specific details lost during pretraining due to distributed-attention. 

Attention map visualizations after SBoRA fine-tuning (Fig. \ref{fig:attn-map-omni-sbora}) confirm this recovery. For audio, the high attention area after the fine-tuning (Fig. \ref{fig:attn-map-omni-audio-sbora}) become less spread out compared to pretrained state (Fig. \ref{fig:attn-map-omni-audio}). The attention areas emerge as straighter and thicker stripes more akin to the expert model, indicating partial recovery of audio-specialized attention. For text, comparisons between pretraining (Fig. \ref{fig:attn-map-omni-text}) and after fine-tuning (Fig. \ref{fig:attn-map-omni-text-sbora}) reveal that the block-grid effect diminishes, while high-intensity sparse straight lines intensify, bringing the patterns closer to those of the expert text model after fine-tuning (Fig.\ref{fig:attn-map-expert-text-sbora}). This performance parity aligns with our hypothesis that the unified model, acting as a lossy universal compressor, preserves foundational features that can be efficiently refined for downstream tasks using parameter-efficient methods. 

\begin{figure}[t]
\centering
\includegraphics[width=0.9\linewidth]{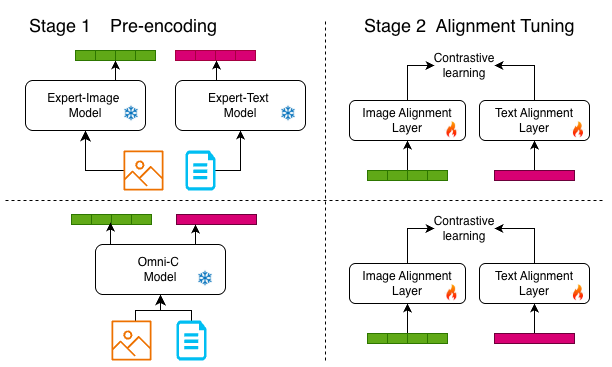}
\caption{SAIL-based \cite{zhang2025assessing} alignment workflow. Features are extracted from the image-text pairs in stage 1. A linear probe is trained in the stage 2 for modality alignment. The same workflow is applied for audio-text alignment.}
\label{fig:alignment-workflow}
\end{figure}

\begin{table*}[t]
    \caption{Evaluation for CLIP-style zero-shot image classification for image-text aligned ViT-Base-32 model. IParam represents the inference parameter. I, A, and T denote Image, Audio, and Text, respectively.}
    \centering
    \resizebox{2\columnwidth}{!}{
        \begin{tabular}{l c c c c c c c c c c c c c c}
            \hline
            Model & IParam (M) & Cars & DTD & EuroSAT & GTSRB  & KITTI & MNIST & RESISC45 & SUN397 & SVHM & \textbf{Avg Acc} \\
            \hline
            \cellcolor{lightgray}Expert-Image-Text \cite{zhang2025assessing}  & \cellcolor{lightgray}196.4 & \cellcolor{lightgray}0.60 & \cellcolor{lightgray}\textbf{5.02} & \cellcolor{lightgray}16.03 & \cellcolor{lightgray}\textbf{2.86} & \cellcolor{lightgray}41.26 & \cellcolor{lightgray}\textbf{13.22} & \cellcolor{lightgray}\textbf{8.82} & \cellcolor{lightgray}\textbf{13.97} & \cellcolor{lightgray}7.83 & \cellcolor{lightgray}12.17 \\
            \hline
            Omni-C (I \& T) & 111.1 & \textbf{0.66} & 4.91 & \textbf{17.39} & 2.36 & 53.42 & 3.70 & 8.89 & 11.15 & 11.25 & 12.63 \\
            \hline
            Omni-C (I \& A \& T) & 111.9 & 0.42 & 1.23 & 16.66 & 1.75 & \textbf{58.35} & 12.72 & 7.70 & 9.04 & \textbf{16.30} & \textbf{13.79} \\
            \hline

        \end{tabular}
    }
    \label{tab:zero-shot-alignment-image}
\end{table*}

\begin{table}[t]
    \centering
    \caption{Evaluation for CLIP-style zero-shot audio classification for audio-text aligned ViT-Base-32 model. IParam represents the inference parameter. I, A, and T denote Image, Audio, and Text, respectively.}
    \resizebox{1\columnwidth}{!}{
    \begin{tabular}{l c c c c c c}
        \hline
         \textbf{Model}  & \textbf{IParam(M)} & \textbf{VGGSound} & \textbf{EPIC-Sound} & \textbf{SpeechCommand} & \textbf{Nsynth} &  \textbf{AvgAcc} \\
        \hline
        \cellcolor{lightgray}Expert-Audio-Text & \cellcolor{lightgray}194.8 & \cellcolor{lightgray}\textbf{2.27} & \cellcolor{lightgray}2.37 & \cellcolor{lightgray}3.33 & \cellcolor{lightgray}\textbf{9.80} & \cellcolor{lightgray}4.44 \\
         \hline
        
        Omni-C (A \& T) & 109.5 & 1.37 & 1.41 & 2.31 & 3.65 & 2.18 \\
        \hline

        Omni-C (I \& A \& T) & 111.9 & 0.88 & \textbf{4.39} & \textbf{3.44} & 9.36 & \textbf{4.51} \\
        \hline
        \end{tabular}
    }
    \label{tab:zero-shot-alignment-audio}
\end{table}

\subsection{Cross-Model Alignment and Generalization}
\label{sec:cross-model-align}
To evaluate the cross-modal generalization capabilities of our pretrained Omni-C (Image \& Audio \& Text) model, we conduct alignment experiments for image-text and audio-text tasks using an efficient protocol called SAIL \cite{zhang2025assessing}.

\noindent\textbf{SAIL Alignment Setup.} Following the alignment approach, SAIL,  proposed in \cite{zhang2025assessing}, we perform cross-modal alignment by attaching a pair of trainable linear projection layers—one per modality—on top of the respective frozen unimodal encoders. The modality-specific backbones are kept frozen throughout training, allowing gradients to update solely the parameters of these lightweight linear layers. We perform similar operation using Omni-C model as well. The difference between using SAIL with modality-sepcific model and with Omni-C model is shown in the Figure \ref{fig:alignment-workflow}. To achieve alignment, we employ the CLIP-style symmetric InfoNCE loss \cite{zhang2025assessing},\cite{radford2021learning}. The loss pulls positive paired cross-modal samples closer while repelling negatives within each batch. 
 
\noindent\textbf{Training Details.} For image-text alignment, we use paired data from the Conceptual Captions 3M (CC3M) dataset \cite{changpinyo2021conceptual}. For audio-text alignment, we use approximately 50K paired audio-caption samples from the AudioCaps dataset \cite{audiocaps}. To ensure efficient training, we pre-extract and cache all frozen backbone features offline as mentioned above. Alignment is then performed on these precomputed features using 1 RTX 3090 GPUs with batch size of 1024, AdamW optimizer, learning rate of 1e-3 with minimum 1e-4, weight decay of 0.1, and 10 warmup epochs. Image-text alignment is trained for 100 epochs. Audio-text alignment is trained for 200 epochs to allow sufficient convergence given the smaller paired dataset size. The setup remains highly resource-efficient and consumes approximately 2GB of GPU memory.

\noindent\textbf{Evaluation Protocol.} Following alignment training, we evaluate zero-shot classification performance on the downstream tasks listed in Table~\ref{tab:downstream-datasets}. We adopt the standard CLIP-style zero-shot protocol \cite{radford2021learning}. For each input sample (image or audio), the modality-specific encoder followed by the learned linear projection produces an embedding in the joint space. This projected embedding is then compared, via cosine similarity, to the text embeddings of class-specific prompts (e.g., “a photo of a [class]”). The predicted class is the one with the highest similarity score.

\noindent\textbf{Results.} 
Table~\ref{tab:zero-shot-alignment-image} presents zero-shot classification performance on image-based downstream tasks following image–text alignment training. The aligned Omni-C model achieves an average top-1 accuracy of 13.79\%, outperforming the expert image–text baseline (12.17\%) by a margin of 1.62 \%. This improvement demonstrates enhanced generalization in the unified multimodal model, likely attributable to the transfer of knowledge across modalities during alignment.In contrast, Table~\ref{tab:zero-shot-alignment-audio} reports zero-shot classification performance on audio downstream tasks after audio–text alignment. The aligned Omni-C model attains an average top-1 accuracy of 4.51\%, compared to 4.34\% for the expert audio–text baseline—a modest gain of 0.17\%. These results indicate near-parity between the unified Omni-C model and the specialized audio–text expert on zero-shot audio classification, suggesting effective preservation of unimodal performance within the shared architecture. 

\noindent\textbf{Inference Memory Usage.} A key practical advantage of the aligned Omni-C model is its significantly reduced parameter count during inference. The unified Omni-C model requires only 111.9M parameters, compared to approximately 196.4M parameters for deploying two separate expert models (image and text) or 194.8M for audio and text. This substantial parameter saving translates to lower memory usage during inference and making Omni-C particularly suitable for memory-constrained edge devices. With a compact model footprint and support for sequential modality processing, it enables low-memory and low-power deployment without the need for parallel expert loading.

These findings highlight the core advantages of our approach. A single dense encoder naturally acts as a lossy universal compressor for different modalities, and it can adapt to the multimodal alignment processes without performance degradation compared to multi-expert model alignment. At the same time, Omni-C reduces memory and deployment complexity for alignment training and inference.

\section{Ablation Studies}
\label{sec:ablation_studies}

\begin{figure}
   \centering
   \includegraphics[width=\linewidth]{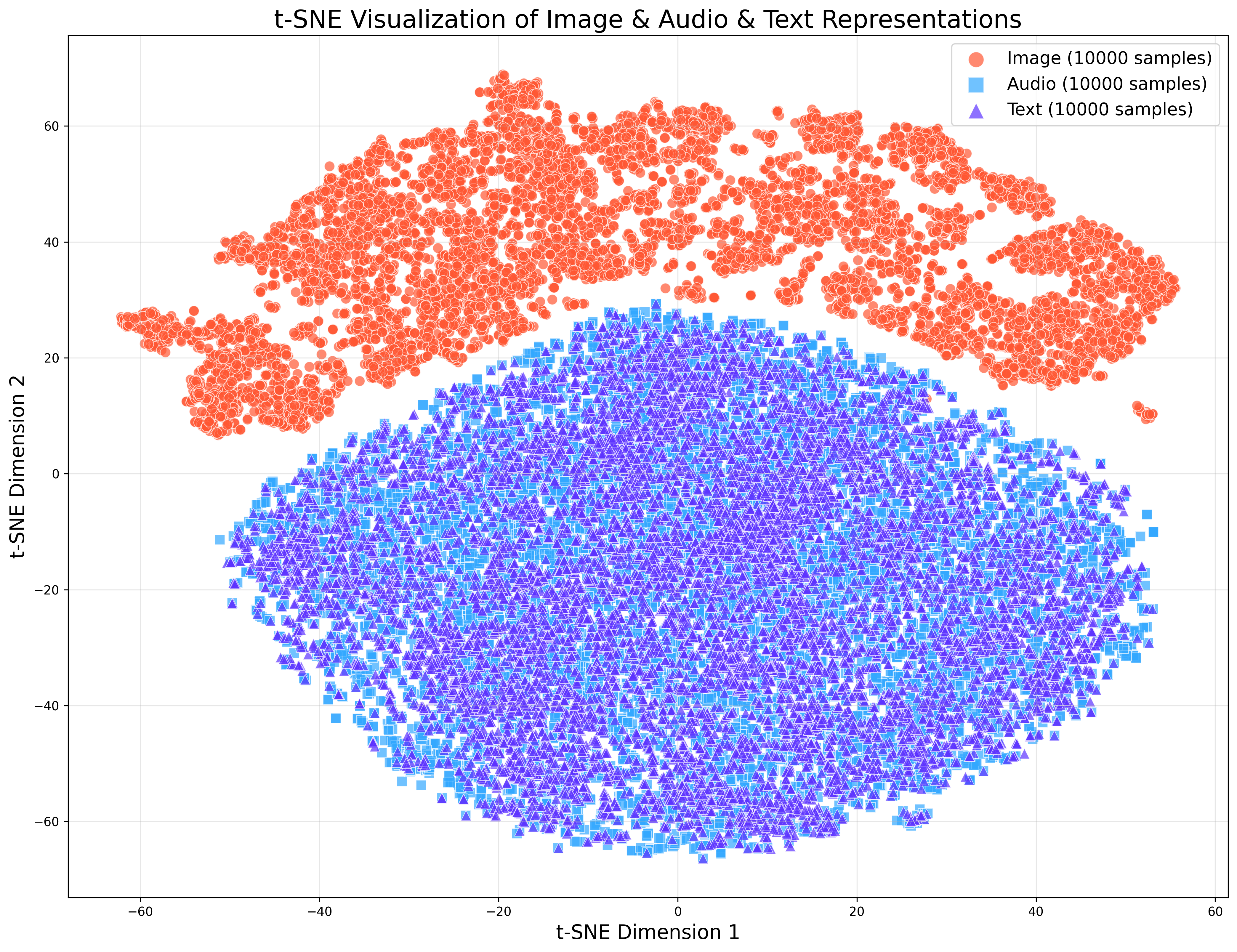} 
   \caption{t-SNE visualization of image, audio, and text embeddings of Omni-C model with share projector head. Samples from ImageNet-1K (red), AudioSet (blue) and English Wikipedia (green). With a shared projector, the embeddings of audio (blue) and text (green) show significant overlapping and mixture in the shared embedding space}
\label{fig:tsne_image_audio_text_unified_share_projector}
\end{figure}

\begin{table*}[t]
    \caption{Comparison of \textbf{zero shot, linear probe and SBORA fine-tuning} performance on downstream \textbf{image classification tasks} for pretrained ViT-Base-32 models under different projector settings (shared vs. separate projectors). Here we use unified Omni-C model with image, audio and text modalities. SP represent share projector and MP represent share projector.}
    \centering
    \resizebox{2\columnwidth}{!}{
        \begin{tabular}{l c c c c c c c c c c c}
            \hline
            Evaluation & Model  & Cars & DTD & EuroSAT & GTSRB  & KITTI & MNIST & RESISC45 & SUN397 & SVHM & \textbf{Avg Acc} \\
            \hline
            \multirow{2}{*}{ZS} & Omni-C-SP & \textbf{2.58} & 32.37 & \textbf{75.15} & \textbf{24.19} & 52.88 & 47.07 & 42.78 & 19.79 & \textbf{22.40} & 35.46\\
            
            & Omni-C-MP  & 2.22 & \textbf{33.37} & 74.97 & 21.40 & \textbf{57.18} & \textbf{47.84} & \textbf{42.93} & \textbf{20.01} & 21.80 & \textbf{35.74}\\
             \hline

            \multirow{2}{*}{LP} & Omni-C-SP & 5.95 & 36.16 & 89.40 & 69.23 & 90.76 & 87.37 & 64.13 & 43.34 & 41.66 & 58.66\\
            
            & Omni-C-MP  & \textbf{8.75} & \textbf{41.18} & \textbf{91.28} & \textbf{76.40} & \textbf{91.41} & \textbf{92.87} & \textbf{72.07} & \textbf{49.03} & \textbf{49.19} & \textbf{63.57}\\
            \hline

            \multirow{2}{*}{SBORA FT} & Omni-C-SP  & \textbf{46.07} & 53.20 & 98.12 & 99.76 & 98.02 & 99.59 & 87.87 & 59.30 & 94.43 & 81.81 \\
            
            & Omni-C-MP  & 45.99 & \textbf{53.36} & \textbf{98.22} & \textbf{99.88} & \textbf{98.41} & \textbf{99.54} & \textbf{88.30} & \textbf{60.31} & \textbf{94.60} & \textbf{82.06} \\
            \hline
            
        \end{tabular}
    }
    \label{tab:image-sp-vs-mp}
\end{table*}

\begin{table}[t]
    \centering
    \caption{Comparison of \textbf{zero shot, linear probe and SBORA fine-tuning} performance on downstream \textbf{audio classification tasks} for pretrained ViT-Base-32 models under different projector settings (shared vs. separate projectors). Here we use unified Omni-C model with image, audio and text modalities. SP represent share projector and MP represent share projector}
    \resizebox{1\columnwidth}{!}{
    \begin{tabular}{l c c c c c c}
        \hline
        \textbf{Evaluation} & \textbf{Model}  & \textbf{VGGSound} & \textbf{EPIC-Sound} & \textbf{SpeechCommand} & \textbf{Nsynth} &  \textbf{AvgAcc} \\
        \hline
        \multirow{2}{*}{ZS} & Omni-C-SP &  2.36 & 4.61 & 9.00 & \textbf{23.43} & 9.85 \\
        
        & Omni-C-MP & \textbf{2.63} & \textbf{4.61} & \textbf{9.00} & 13.43 & \textbf{9.91} \\
        \hline

        \multirow{2}{*}{LP} &  Omni-C-SP &  16.41 & \textbf{32.86} & 34.97 & 53.99 & 34.55 \\
        
        & Omni-C-MP & \textbf{17.12} & 32.68 & \textbf{34.98} & \textbf{54.64} & \textbf{34.85} \\
        \hline

        \multirow{2}{*}{SBORA FT} & Omni-C-SP &  32.63 & 39.07 & 86.64 & 71.73 & 57.51 \\
        
        & Omni-C-MP &  \textbf{33.39} & \textbf{39.07} & \textbf{87.61} & \textbf{72.47} & \textbf{58.13} \\
        \hline

        \end{tabular}
    }
    \label{tab:audio-sp-vs-mp}
\end{table}

\begin{table}[t]
    \centering
    \caption{Comparison of \textbf{zero shot, linear probe and SBORA fine-tuning} performance on downstream \textbf{text classification tasks} for pretrained ViT-Base-32 models under different projector settings (shared vs. separate projectors). Here we use unified Omni-C model with image, audio and text modalities. SP represent share projector and MP represent share projector}
    \resizebox{1\columnwidth}{!}{
    \begin{tabular}{l c c c c c c}
        \hline
        \textbf{Evaluation} & \textbf{Model}  & \textbf{AGNEWS} & \textbf{NEWSGROUPS} & \textbf{IMDB} & \textbf{CARER} &  \textbf{AvgAcc} \\
        \hline
        \multirow{2}{*}{ZS} & Omni-C-SP &  55.67 & \textbf{12.98} & \textbf{57.23} & \textbf{19.22} & \textbf{36.27} \\
        
        & Omni-C-MP & \textbf{56.08} & 12.82 & 52.85 & 17.08 & 34.70 \\
        \hline

        \multirow{2}{*}{LP} & Omni-C-SP &  87.15 & 43.83 & \textbf{68.67} & 42.84 & 60.62 \\
        
        & Omni-C-MP & \textbf{88.22} & \textbf{45.36} & 68.13 & \textbf{43.68} & \textbf{61.34} \\
        \hline

        \multirow{2}{*}{SBORA FT} & Omni-C-SP &  93.53 & 59.51 & 82.35 & 88.55 & 80.98 \\
        
        & Omni-C-MP & \textbf{93.72} & \textbf{60.25} & \textbf{82.73} & \textbf{90.24} & \textbf{81.73} \\
        \hline

        \end{tabular}
    }
    \label{tab:text-sp-vs-mp}
\end{table}

\subsection{Separate projectors vs. Share projector}
\label{sec:mp-vs-sp}
To investigate the importance of modality-specific projectors in preserving distinct feature subspaces during joint pretraining, we conduct an ablation study comparing our main approach (using separate modality-specific projectors) against a single shared projector (allowing mixed-modality projection) that maps the global CLS token representations from all modalities into a common lower-dimensional space. The t-SNE visualizations in Fig. \ref{fig:tsne_image_audio_text_unified_seperate_projector} (separate projectors) and Fig. \ref{fig:tsne_image_audio_text_unified_share_projector} (shared projector) clearly illustrates the impact on embedding space organization. With separate projectors (Fig. \ref{fig:tsne_image_audio_text_unified_seperate_projector}), the embeddings form well-defined, non-overlapping clusters for images, audio, and text, demonstrating effective preservation of distinct feature subspaces despite pretraining a shared backbone  on unaligned heterogenous data. In contrast, the shared projector variant (see Fig. \ref{fig:tsne_image_audio_text_unified_share_projector}) exhibits significant mixing, particularly between text and audio embeddings, with no boundaries and overlapping regions. This degradation highlights the importance of dedicated subspaces for each modality, as forcing heterogeneous inputs into a single projected space leads to interference and reduced distinctiveness.

Quantitative results further corroborate these observations. In zero-shot evaluation Tables \ref{tab:image-sp-vs-mp}, \ref{tab:audio-sp-vs-mp}, and \ref{tab:text-sp-vs-mp} (row1 to 2), the separate-projector setup achieves average top-1 accuracies of 35.74\% (images), 9.91\% (audio), and 36.27\% (text), compared to 35.46\%, 9.85\%, and 34.70\% for the shared-projector variant, respectively. Similarly, linear probing (Tables \ref{tab:image-sp-vs-mp}, \ref{tab:audio-sp-vs-mp}, and \ref{tab:text-sp-vs-mp} row3 to 4) yields 63.57\% (images), 34.85\% (audio), and 61.34\% (text) for separate projectors, versus 58.66\%, 34.55\%, and 60.62\% for shared projectors. In SBoRA fine-tuning (Tables \ref{tab:image-sp-vs-mp}, \ref{tab:audio-sp-vs-mp}, and \ref{tab:text-sp-vs-mp} row 5 to 6), the separate-projector results are 82.06\% (images), 58.13\% (audio), and 81.73\% (text), compared to 81.81\%, 57.51\%, and 80.98\% for shared projectors. These consistent improvements across all evaluation protocols, particularly pronounced in text due to both its structural dissimilarity and the observed mixing of text and audio embeddings, confirm that modality-specific projectors are essential for maintaining robust and transferable representations, when ptretaining a shared backbone on heterogeneous data.

\begin{figure}
   \centering
   \includegraphics[width=\linewidth]{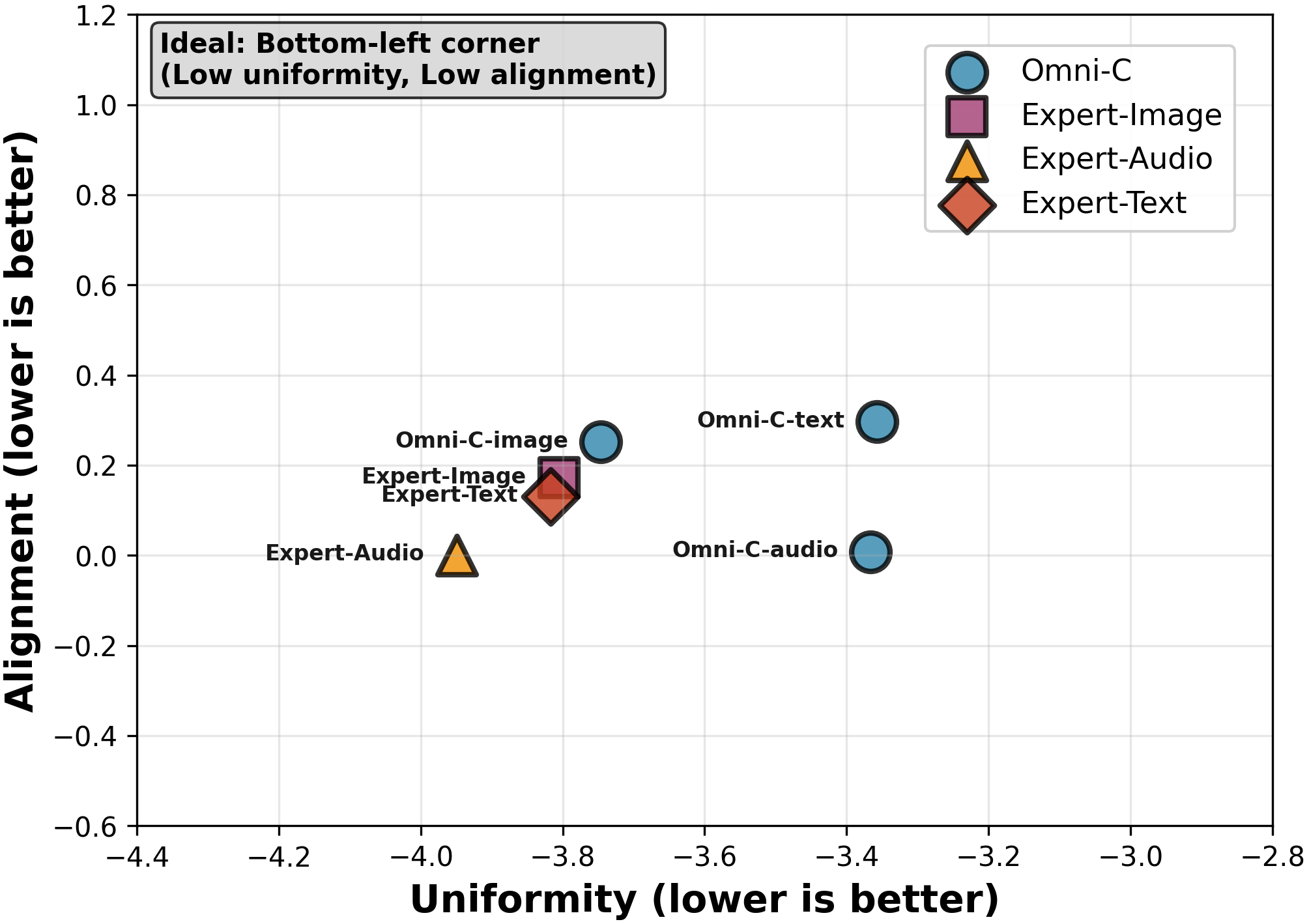} 
   \caption{Uniformity vs. alignment scatter plot of embeddings from the unified Omni-C model (blue circles) and modality-specific expert models (pink square for image, orange triangle for audio, red diamond for text), computed using the SimCSE \cite{gao2021simcse} evaluation protocol. Lower values on both axes are better (ideal region: bottom-left corner, indicating low uniformity and low alignment).}
   \label{fig:alignment_uniformity_expert_vs_omnic}
\end{figure}

\subsection{Alignment and uniformity analysis}
The uniformity vs. alignment scatter plot (Fig. \ref{fig:alignment_uniformity_expert_vs_omnic}), computed following the SimCSE \cite{gao2021simcse} evaluation protocol, provides quantitative evidence of the embedding quality in the unified Omni-C model compared to the modality-specific experts. Alignment (lower values indicate better closeness of positive pairs within each modality) for Omni-C is very close to that of the experts: 0.252 (image), 0.008 (audio), and 0.296 (text), versus 0.172, 0.001, and 0.131 for the respective experts. This demonstrates that the unified model effectively preserves within-modality invariance through unimodal contrastive pretraining, despite dense parameter sharing across heterogeneous modalities. Uniformity (lower values indicate better intra-modality spread), on the other hand, is slightly higher for Omni-C (more concentrated distributions within each modality) compared to the experts (e.g., -3.366 vs. -3.949 for audio), reflecting the expected regularization penalty of joint training. This modest increase is acceptable given the substantial efficiency gains of a single backbone. Combined with the clear, non-overlapping modality clusters observed in the t-SNE visualization (Fig. \ref{fig:tsne_image_audio_text_unified_seperate_projector}), these metrics confirm that the Omni-C model achieves a balanced representation space: strong intra-modality similarity and clear inter-modality distinction in a single efficient backbone, supporting our hypothesis of internal neural pathways that enable effective modality separation without explicit cross-modal supervision.

\section{Conclusion}
\label{sec:conclusion}
In this paper, we introduced Omni-C, a single dense Transformer-based encoder that learns competitive shared representations across images, audio, and text via unimodal contrastive pretraining on large-scale unaligned data. By maximizing parameter sharing in the backbone and using lightweight modality-specific projection heads, Omni-C effectively mitigates inter-modality conflicts without requiring Mixture-of-Experts, paired supervision, or routing mechanisms. This design enables highly efficient deployment on memory-constrained edge devices with low-memory inference, while achieving nearly $3\times$ parameter savings for the three modalities (images, audio, and text) compared to the deployment of separate expert models per modality. Experimental results demonstrate that Omni-C performs competitively to unimodal expert models in both unimodal and cross-modal settings. In the future, this work can be extended to additional modalities such as video, sensor data like IMU, thermal imaging or depth maps, to further investigate the capability and limits of a single shared encoder in handling diverse heterogeneous inputs.

\bibliographystyle{IEEEtran}
\bibliography{egbib}

\newpage

 




\vfill

\end{document}